\newcommand{\psSurveyArea}{296}

\newcommand{\LCDM}{$\Lambda$CDM}

\newcommand{\be}{\begin{equation}}
\newcommand{\ee}{\end{equation}}
\newcommand{\ba}{\begin{eqnarray}}
\newcommand{\ea}{\end{eqnarray}}

\newcommand{\wmap}    {{\sl WMAP}}
\newcommand{\act}    {ACT}
\newcommand{\neff}  {$N_{\rm eff}$}
\newcommand{\arone}{148\,GHz}
\newcommand{\artwo}{218\,GHz}
\newcommand{\degree}{\ensuremath{^{\circ}}}
\newcommand{\etal}{et al.\,}
\newcommand{\mat}[1]{\ensuremath{\mathbf #1}}

\documentclass[numberedappendix,letter]{emulateapj}
\usepackage{graphicx}        
\usepackage{bm}              
\usepackage{natbib}
\usepackage{mathrsfs}

\shorttitle{\act\ 2008 Parameters}
\shortauthors{J.~Dunkley \etal}

\begin{document}

\title{The Atacama Cosmology Telescope: Cosmological Parameters from the 2008 Power Spectra}
\author{
J.~Dunkley\altaffilmark{1,2,3},
R.~Hlozek\altaffilmark{1},
J.~Sievers\altaffilmark{4},
V.~Acquaviva\altaffilmark{5,3},
P.~A.~R.~Ade\altaffilmark{6},
P.~Aguirre\altaffilmark{7},
M.~Amiri\altaffilmark{8},
J.~W.~Appel\altaffilmark{2},
L.~F.~Barrientos\altaffilmark{7},
E.~S.~Battistelli\altaffilmark{9,8},
J.~R.~Bond\altaffilmark{4},
B.~Brown\altaffilmark{10},
B.~Burger\altaffilmark{8},
J.~Chervenak\altaffilmark{11},
S.~Das\altaffilmark{12,2,3},
M.~J.~Devlin\altaffilmark{13},
S.~R.~Dicker\altaffilmark{13},
W.~Bertrand~Doriese\altaffilmark{14},
R.~D\"{u}nner\altaffilmark{7},
T.~Essinger-Hileman\altaffilmark{2},
R.~P.~Fisher\altaffilmark{2},
J.~W.~Fowler\altaffilmark{14,2},
A.~Hajian\altaffilmark{4,3,2},
M.~Halpern\altaffilmark{8},
M.~Hasselfield\altaffilmark{8},
C.~Hern\'andez-Monteagudo\altaffilmark{15},
G.~C.~Hilton\altaffilmark{14},
M.~Hilton\altaffilmark{16,17},
A.~D.~Hincks\altaffilmark{2},
K.~M.~Huffenberger\altaffilmark{18},
D.~H.~Hughes\altaffilmark{19},
J.~P.~Hughes\altaffilmark{5},
L.~Infante\altaffilmark{7},
K.~D.~Irwin\altaffilmark{14},
J.~B.~Juin\altaffilmark{7},
M.~Kaul\altaffilmark{13},
J.~Klein\altaffilmark{13},
A.~Kosowsky\altaffilmark{10},
J.~M~Lau\altaffilmark{20,21,2},
M.~Limon\altaffilmark{22,13,2},
Y-T.~Lin\altaffilmark{23,3,7},
R.~H.~Lupton\altaffilmark{3},
T.~A.~Marriage\altaffilmark{24,3},
D.~Marsden\altaffilmark{13},
P.~Mauskopf\altaffilmark{6},
F.~Menanteau\altaffilmark{5},
K.~Moodley\altaffilmark{16,17},
H.~Moseley\altaffilmark{11},
C.~B~Netterfield\altaffilmark{25},
M.~D.~Niemack\altaffilmark{14,2},
M.~R.~Nolta\altaffilmark{4},
L.~A.~Page\altaffilmark{2},
L.~Parker\altaffilmark{2},
B.~Partridge\altaffilmark{26},
B.~Reid\altaffilmark{27,2},
N.~Sehgal\altaffilmark{20},
B.~Sherwin\altaffilmark{2},
D.~N.~Spergel\altaffilmark{3},
S.~T.~Staggs\altaffilmark{2},
D.~S.~Swetz\altaffilmark{13,14},
E.~R.~Switzer\altaffilmark{28,2},
R.~Thornton\altaffilmark{13,29},
H.~Trac\altaffilmark{30,31},
C.~Tucker\altaffilmark{6},
R.~Warne\altaffilmark{16},
E.~Wollack\altaffilmark{11},
Y.~Zhao\altaffilmark{2}
}
\altaffiltext{1}{Sub-department of Astrophysics, University of Oxford, Denys Wilkinson Building, Keble Road, Oxford OX1 3RH, UK}
\altaffiltext{2}{Joseph Henry Laboratories of Physics, Jadwin Hall,
Princeton University, Princeton, NJ, USA 08544}
\altaffiltext{3}{Department of Astrophysical Sciences, Peyton Hall, 
Princeton University, Princeton, NJ USA 08544}
\altaffiltext{4}{Canadian Institute for Theoretical Astrophysics, University of
Toronto, Toronto, ON, Canada M5S 3H8}
\altaffiltext{5}{Department of Physics and Astronomy, Rutgers, 
The State University of New Jersey, Piscataway, NJ USA 08854-8019}
\altaffiltext{6}{School of Physics and Astronomy, Cardiff University, The Parade, 
Cardiff, Wales, UK CF24 3AA}
\altaffiltext{7}{Departamento de Astronom{\'{i}}a y Astrof{\'{i}}sica, 
Facultad de F{\'{i}}sica, Pontific\'{i}a Universidad Cat\'{o}lica,
Casilla 306, Santiago 22, Chile}
\altaffiltext{8}{Department of Physics and Astronomy, University of
British Columbia, Vancouver, BC, Canada V6T 1Z4}
\altaffiltext{9}{Department of Physics, University of Rome ``La Sapienza'', 
Piazzale Aldo Moro 5, I-00185 Rome, Italy}
\altaffiltext{10}{Department of Physics and Astronomy, University of Pittsburgh, 
Pittsburgh, PA, USA 15260}
\altaffiltext{11}{Code 553/665, NASA/Goddard Space Flight Center,
Greenbelt, MD, USA 20771}
\altaffiltext{12}{Berkeley Center for Cosmological Physics, LBL and
Department of Physics, University of California, Berkeley, CA, USA 94720}
\altaffiltext{13}{Department of Physics and Astronomy, University of
Pennsylvania, 209 South 33rd Street, Philadelphia, PA, USA 19104}
\altaffiltext{14}{NIST Quantum Devices Group, 325
Broadway Mailcode 817.03, Boulder, CO, USA 80305}
\altaffiltext{15}{Max Planck Institut f\"ur Astrophysik, Postfach 1317, 
D-85741 Garching bei M\"unchen, Germany}
\altaffiltext{16}{Astrophysics and Cosmology Research Unit, School of
Mathematical Sciences, University of KwaZulu-Natal, Durban, 4041,
South Africa}
\altaffiltext{17}{Centre for High Performance Computing, CSIR Campus, 15 Lower
Hope St., Rosebank, Cape Town, South Africa}
\altaffiltext{18}{Department of Physics, University of Miami, Coral Gables, 
FL, USA 33124}
\altaffiltext{19}{Instituto Nacional de Astrof\'isica, \'Optica y 
Electr\'onica (INAOE), Tonantzintla, Puebla, Mexico}
\altaffiltext{20}{Kavli Institute for Particle Astrophysics and Cosmology, Stanford
University, Stanford, CA, USA 94305-4085}
\altaffiltext{21}{Department of Physics, Stanford University, Stanford, CA, 
USA 94305-4085}
\altaffiltext{22}{Columbia Astrophysics Laboratory, 550 W. 120th St. Mail Code 5247,
New York, NY USA 10027}
\altaffiltext{23}{Institute for the Physics and Mathematics of the Universe, 
The University of Tokyo, Kashiwa, Chiba 277-8568, Japan}
\altaffiltext{24}{Dept. of Physics and Astronomy, The Johns Hopkins University, 3400 N. Charles St., Baltimore, MD 21218-2686}
\altaffiltext{25}{Department of Physics, University of Toronto, 
60 St. George Street, Toronto, ON, Canada M5S 1A7}
\altaffiltext{26}{Department of Physics and Astronomy, Haverford College,
Haverford, PA, USA 19041}
\altaffiltext{27}{Institut de Ciencies del Cosmos (ICC), University of
Barcelona, Barcelona 08028, Spain}
\altaffiltext{28}{Kavli Institute for Cosmological Physics, 
5620 South Ellis Ave., Chicago, IL, USA 60637}
\altaffiltext{29}{Department of Physics , West Chester University 
of Pennsylvania, West Chester, PA, USA 19383}
\altaffiltext{30}{Department of Physics, Carnegie Mellon University, Pittsburgh, PA 15213}
\altaffiltext{31}{Harvard-Smithsonian Center for Astrophysics, 
Harvard University, Cambridge, MA, USA 02138}

\begin{abstract}

 We present cosmological parameters derived from the angular power spectrum 
of the cosmic microwave background (CMB) radiation observed at \arone\ and \artwo\ over \psSurveyArea\,deg$^2$ with the Atacama Cosmology Telescope (\act) during its 2008 season.  \act\ measures fluctuations at scales $500<\ell<10000$.
We fit a model for the lensed CMB, Sunyaev-Zel'dovich (SZ), and foreground contribution to the \arone\ and \artwo\ power spectra, including thermal and kinetic SZ, Poisson power from radio and infrared point sources, and clustered power from infrared point sources. At $\ell=3000$, about half the power at \arone\ comes from primary CMB after masking bright radio sources. The power from thermal and kinetic SZ is estimated to be ${\cal B}_{3000}=6.8\pm2.9~\mu {\rm K}^2$, where  ${\cal B}_\ell \equiv \ell(\ell+1){C}_\ell/2\pi$. The IR Poisson power at \arone\ is ${\cal B}_{3000} = 7.8\pm0.7~\mu {\rm K}^2$ ($C_\ell= 5.5\pm0.5~{\rm nK}^2$), and a clustered IR component is required with ${\cal B}_{3000} = 4.6\pm0.9~\mu {\rm K}^2$, assuming an analytic model for its power spectrum shape. At \artwo\ only about 15\% of the power, approximately $27~\mu {\rm K}^2$, is CMB anisotropy at $\ell=3000$. The remaining 85\% is attributed to IR sources (approximately 50\% Poisson and 35\% clustered), with spectral index $\alpha=3.69\pm0.14$ for flux scaling as $S(\nu) \propto \nu^\alpha$. We estimate primary cosmological parameters from the less contaminated \arone\ spectrum, marginalizing over SZ and source power.  The \LCDM\ cosmological model is a good fit to the data ($\chi^2/{\rm dof}=29/46$), and \LCDM\ parameters estimated from \act+\wmap\ are consistent with the 7-year \wmap\ limits, with scale invariant $n_s=1$ excluded at 99.7\% CL (3$\sigma$). A model with no CMB lensing is disfavored at  2.8$\sigma$. By measuring the third to seventh acoustic peaks, and probing the Silk damping regime, the ACT data improve limits on cosmological parameters that affect the small-scale CMB power. The \act\ data  combined with \wmap\ give a 6$\sigma$ detection of primordial helium, with $Y_P=0.313\pm0.044$, and a $4\sigma$ detection of relativistic species, assumed to be neutrinos,  with $N_{\rm eff}=5.3\pm1.3$ ($4.6\pm0.8$ with BAO+$H_0$ data). From the CMB alone the running of the spectral index is constrained to be $dn_s/d \ln k=-0.034\pm0.018$, the limit on the tensor-to-scalar ratio is $r<0.25$ (95\% CL), and the possible contribution of Nambu cosmic strings to the power spectrum is constrained to string tension $G\mu<1.6 \times 10^{-7}$ (95\% CL). 
\end{abstract}


\section{Introduction}
\label{Sec:intro}

\setcounter{footnote}{0}

Measurements of anisotropies in the cosmic microwave background (CMB) have 
played a central role in the development of the current concordance 
cosmological model \citep[e.g.,][]{smoot/etal:1992,miller/etal:1999,deBernardis/etal:2000,hanany/etal:2000,spergel/etal:2003}.
The \LCDM\ model describes a flat universe with 5\% normal matter, 23\% dark matter, 72\% dark energy, and power-law Gaussian primordial fluctuations consistent with simple inflationary models \citep[see e.g.,][]{komatsu/etal:prep}. Its parameters have been measured to a few percent level accuracy, using CMB data from the \wmap\ satellite and higher resolution experiments, combined with observations of large scale structure and the local expansion rate \citep{brown/etal:2009,riess/etal:2009,reid/etal:2010,percival/etal:2010,larson/etal:prep,komatsu/etal:prep}. The model fits a range of recent astronomical data including Type Ia supernova \citep{hicken/etal:2009,kessler/etal:2009}, galaxy cluster measurements \citep{vikhlinin/etal:2009,mantz/etal:2010,rozo/etal:2010} and gravitational lensing observations \citep{massey/etal:2007,fu/etal:2008,schrabback/etal:2010,suyu/etal:2010}.

The \wmap\ satellite has measured the CMB over the full sky down to 0.2\degree\ resolution. Measurements at higher resolution have been made by a set of complementary balloon and ground-based experiments \citep[e.g.,][]{jones/etal:2006,brown/etal:2009, reichardt/etal:2009,sievers/etal:2009}. The Atacama Cosmology Telescope (ACT) now measures fluctuations on scales from $\simeq$0.4\degree\ to an arcminute. 
The signal observed in this angular range 
is composed of the damped acoustic peaks of the 
primordial CMB signal \citep{silk:1968}, subsequently lensed by large-scale structure, as well as point source emission and fluctuations due 
to the Sunyaev-Zel'dovich (SZ) effect~\citep{sunyaev/zeldovich:1970}, in which CMB photons scatter
off electrons in the hot intra-cluster and filamentary inter-galactic media.  Limits on the SZ power spectrum have been reported from the ACBAR, CBI and SZA experiments \citep{reichardt/etal:2009,sievers/etal:2009,sharp/etal:2010}, with a recent detection reported 
by the South Pole Telescope (SPT) \citep{lueker/etal:prep}.
The first power spectrum measurement from ACT \citep{fowler/etal:prep} provided a limit on the SZ power spectrum, as well as on the point source contribution.

In this paper we present cosmological parameter constraints from power spectra estimated from the \act\ 2008 observing season. We use the  power spectrum to constrain a model for the SZ and point source contribution in the \act\ \arone\ and \artwo\ data. We then combine the \act\ \arone\ data with \wmap\ observations, and additional cosmological distance measures, to constrain the \LCDM\ model and a set of extensions that have particular effects at small scales. 

This is one of a set of papers on the \act\ 2008 data in the southern sky: \citet{swetz/etal:prep} describes the ACT experiment; \citet{dunner/etal:prep} describes the observing strategy and the data; \citet{hajian/etal:prep} describes the calibration to \wmap; \citet{das/etal:prep} presents the power spectra measured at \arone\ and \artwo, and this paper estimates parameters from the power spectrum results. A high-significance SZ galaxy cluster catalog is presented in \citet{marriage/etal:2010b}, with multi-wavelength observations described in \citet{menanteau/etal:prep}  and their cosmological interpretation in \citet{sehgal/etal:2010b}. 
\citet{marriage/etal:2010a} presents the \arone\ point source catalog. \citet{hincks/etal:prep} and \citet{fowler/etal:prep} presented the first maps of clusters and power spectra respectively, from a preliminary version of these maps. Improved map-making and power spectrum estimates, with use of a larger area of sky, now allow us to place new constraints on cosmological models.

The paper is structured as follows. In \S\ref{sec:methods} we describe the \act\ likelihood and parameter estimation methodology. In \S\ref{sec:sz_ptsrc} we present 
results on SZ and point source parameters from the small-scale power spectra. In \S\ref{sec:cosmo_params} we present constraints on a set of cosmological models in combination with other cosmological data, and conclude in \S\ref{sec:discuss}.

\section{Methodology}
\label{sec:methods}

This section describes the methods used to estimate parameters from the \act\ power spectra. The power spectra, estimated from \psSurveyArea\,deg$^2$ of sky observed in 2008, are described in \cite{das/etal:prep}, and details of the experiment, data reduction and map-making are described in \cite{swetz/etal:prep} and \citet{dunner/etal:prep}.  We will estimate two sets of parameters: `primary' and `secondary'. Primary parameters describe the cosmological model from which a theoretical primary CMB power spectrum can be computed. Secondary parameters describe the additional power from SZ fluctuations and foregrounds. We construct an \act\ likelihood function that returns the probability of the ACT data given some theoretical CMB power spectrum and a set of secondary parameters. Primary and secondary parameters are then estimated from \act\ and additional datasets using Markov Chain Monte Carlo (MCMC) methods. The \act\ likelihood function is described in Section \ref{subsec:act_like}, and the MCMC methods in \ref{subsec:param_methods}.

\subsection{\act\ Likelihood}
\label{subsec:act_like}

For maps of temperature fluctuations at \arone\ and \artwo, three cross-spectra are estimated in bands in the range $500<\ell<10000$ for \arone, and $1500<\ell<10000$ for both the \artwo\ and the $148\times218$~GHz cross-spectrum \citep[see][]{das/etal:prep}. A likelihood function is constructed to estimate parameters from these spectra. The function returns the likelihood of the data, $p(d|C_\ell^{\rm CMB}, \Theta)$, given a theoretical lensed CMB spectrum, $C_\ell^{\rm CMB}$, and a set of secondary parameters describing the additional small-scale power, $\Theta$.  In this analysis we consider two likelihoods: the `148+218' likelihood, which returns the likelihood of all three spectra given a model, and the `148-only' likelihood, which returns the likelihood of just the \arone\ spectrum given a model.

The temperature fluctuations in the ACT maps are expected to be the sum of fluctuations from lensed CMB, thermal and kinetic SZ, radio point sources, infrared point sources, and thermal dust emission from the Galaxy \citep[see e.g.,][]{sehgal/etal:2010a}. 
These vary as functions of frequency. The lensed CMB and the kinetic SZ are blackbody emission, and so are constant as a function of frequency in thermodynamic units. The thermal SZ emission has a known frequency dependence \citep{sunyaev/zeldovich:1970} and has negligible contribution at \artwo. The radio point sources emit synchrotron, and the IR sources emit thermally, so both can be modeled as following power law emission in flux given a narrow enough frequency range. 
For frequency $\nu$ and direction ${\bf\hat{n}}$ the signal in the maps can be modeled in thermal units as
\be
\Delta T(\nu,{\bf\hat{n}}) = \Delta T^{\rm CMB}({\bf\hat{n}}) + \Delta T^{\rm SZ}(\nu,{\bf\hat{n}})+\Delta T^{\rm fg}(\nu,{\bf\hat{n}}),
\label{eqn:map_model}
\ee
with lensed CMB fluctuations $\Delta T^{\rm CMB}({\bf\hat{n}})$. The SZ signal is the sum of thermal and kinetic components
\be
\Delta T^{\rm SZ}(\nu,{\bf\hat{n}}) = \frac{f(\nu)}{f(\nu_0)}\Delta T_0^{\rm tSZ}({\bf\hat{n}}) + \Delta T^{\rm kSZ}({\bf\hat{n}}), 
\ee
where the factor $f(\nu)=2-(x/2)/\tanh(x/2)$, for  $x=h\nu/k_BT_{\rm CMB}$, converts the expected SZ emission from the Rayleigh-Jeans (RJ) limit to thermodynamic units, and $\Delta T_0^{\rm tSZ}$ is the expected signal at frequency $\nu_0$. At 218~GHz there is negligible thermal SZ signal, with $f(218)=0$.
The point source and Galactic components are modeled as
\ba
\Delta T^{\rm fg}(\nu,{\bf\hat{n}})  &=& \frac{g(\nu)}{g(\nu_0)}\left\{\Delta T_0^{\rm IR}\left(\frac{\nu}{\nu_0}\right)^{\alpha_d-2} + \Delta T_0^{\rm rad}\left(\frac{\nu}{\nu_0}\right)^{\alpha_s-2} \right. \nonumber\\
& &\qquad\qquad \left. + \Delta T_0^{\rm Gal} \left(\frac{\nu}{\nu_0}\right)^{\alpha_g-2}\right\},
\ea
assuming that the IR and radio source emission in antenna temperature, $\Delta T_0^{\rm IR, \rm rad}$, scale with global power laws $\alpha_d-2$ and $\alpha_s-2$, respectively, where $\alpha$ is the index in flux units. The factor $g(\nu)=(e^x-1)^2/x^2e^x$ converts from antenna to CMB thermodynamic temperature at frequency $\nu$. The factors are $g=[1.71,3.02]$ for \arone\ and \artwo.  Power law behavior is also assumed for the Galactic dust emission $\Delta T_0^{\rm Gal}$.  This behavior is expected to be a good approximation between \arone\ and \artwo, but breaks down over larger frequency ranges. 

To compute the likelihood one could first estimate the CMB map, by subtracting off the foreground and SZ components. This is commonly done in CMB analyses for subtracting Galactic components \citep[e.g.,][]{bennett/etal:2003}, and has also been used for subtracting the IR point sources \citep{hall/etal:2010,lueker/etal:prep}. However, for the ACT frequencies and noise levels the radio sources cannot be neglected, so a linear combination of the \arone\ and \artwo\ maps will not remove all the source contamination. Instead, we choose to construct a model for the cross power spectra between frequency $\nu_i$ and $\nu_j$, 
\be
C^{ij}_\ell=\left<{\tilde T}^*_\ell(\nu_i){\tilde T}_\ell(\nu_j)\right>,
\ee
where ${\tilde T}_\ell$ is the Fourier transform of $\Delta T({\bf\hat n})$. 
For ACT analysis we use a flat-sky approximation, described in \citet{das/etal:prep}.
The individual components are assumed to be uncorrelated, so the theoretical cross-spectrum ${\cal B}_\ell^{\rm{th,ij}}$ is modeled from Eq.~\ref{eqn:map_model} as
\be
\label{eq:model}
{\cal B}_\ell^{\rm{th,ij}}  
                          = {\cal B}_\ell^{\rm{CMB}} 
+ {\cal B}_\ell^{\rm{tSZ,ij}} 
+ {\cal B}_\ell^{\rm{kSZ,ij}}  
+ {\cal B}_\ell^{\rm{IR,ij}}
+  {\cal B}_\ell^{\rm{rad,ij}}
+ {\cal B}_\ell^{\rm{Gal,ij}},
\label{eqn:spectra_th}
\ee
where ${\cal B}_\ell \equiv \ell(\ell+1){C}_\ell/2\pi$. The first term, ${\cal B}_\ell^{\rm {CMB}}$, is the lensed primary CMB power spectrum and is the same at all frequencies. 
The thermal SZ (tSZ) power is modeled as 
\be
{\cal B}_\ell^{\rm{tSZ,ij}}  = A_{\rm tSZ}\frac{f(\nu_i)}{f(\nu_0)}\frac{f(\nu_j)}{f(\nu_0)}{\cal B}_{0,\ell}^{\rm{tSZ}},
\ee
where ${\cal B}_{0,\ell}^{\rm{tSZ}}$ is a template power spectrum corresponding to the predicted tSZ emission at $\nu_0=$\arone\ for a model with $\sigma_8=0.8$, to be described in Section \ref{subsubsec:sz_temp}, and $A_{\rm tSZ}$ is an amplitude parameter. The kinetic SZ (kSZ) power is modeled as 
\be
{\cal B}_\ell^{\rm{kSZ,ij}}  = A_{\rm kSZ}{\cal B}_{0,\ell}^{\rm{kSZ}},
\ee
where ${\cal B}_{0,\ell}^{\rm kSZ}$ is a template spectrum for the predicted blackbody kSZ emission for a model with $\sigma_8=0.8$, also described in Section \ref{subsubsec:sz_temp}. The total SZ power is then given by
\be
{\cal B}_\ell^{\rm{SZ,ij}}  = A_{\rm tSZ}\frac{f(\nu_i)}{f(\nu_0)}\frac{f(\nu_j)}{f(\nu_0)}{\cal B}_{0,\ell}^{\rm{tSZ}}+A_{\rm kSZ}{\cal B}_{0,\ell}^{\rm{kSZ}}.
\ee

The infrared point sources are expected to be clustered, and their power is modeled as
\ba
&&{\cal B}_\ell^{\rm{IR,ij}} =\nonumber \\
&&\left[ A_d\left(\frac{\ell}{3000}\right)^2 +A_c{\cal B}_{0,\ell}^{\rm clust}\right]\frac{g(\nu_i)}{g(\nu_0)}\frac{g(\nu_j)}{g(\nu_0)}\left(\frac{\nu_i}{\nu_0}\frac{\nu_j}{\nu_0}\right)^{\alpha_d-2}
\ea
where $A_d$ and $A_c$ are the values of ${\cal B}^{\rm IR}_{3000}$ at \arone\ for Poisson and clustered dust terms respectively, assuming a normalized template spectrum ${\cal B}_{0,\ell}^{\rm clust}$. This will be described in Section \ref{subsubsec:cluster_temp}. This model assumes the same spectral index for the clustered and Poisson IR power. 
The radio sources are not expected to be significantly clustered \citep[see, e.g.,][]{sharp/etal:2010,hall/etal:2010}, and so can be described by Poisson scale-free power, with
\be
 {\cal B}_\ell^{\rm{rad,ij}}  = A_s\left(\frac{\ell}{3000}\right)^2\frac{g(\nu_i)}{g(\nu_0)}\frac{g(\nu_j)}{g(\nu_0)}\left(\frac{\nu_i}{\nu_0}\frac{\nu_j}{\nu_0}\right)^{\alpha_s-2},
\ee
with amplitude $A_s$ normalized at $\nu_0=$\arone\ and $\ell=3000$. 

Though we have described the pivot frequency as $\nu_0 =$\arone\ for all components in Eqs.~6-10, in practice we use the band-centers for SZ, radio and dusty sources given in Table 4 of \citet{swetz/etal:prep}. The Galactic emission, ${\cal B}_\ell^{\rm{Gal,ij}}$, is expected to be sub-dominant on ACT scales, as demonstrated in \citet{das/etal:prep} using the FDS dust map \citep{finkbeiner/davis/schlegel:1999} as a Galactic dust template, so is neglected in this analysis. 

Given SZ and clustered source templates, aside from parameters constrained by ${\cal B}_\ell^{\rm CMB}$,  this model has 7 free parameters: five amplitudes $A_{\rm tSZ}$, $A_{\rm kSZ}$, $A_d$, $A_c$, $A_s$, and two spectral indices, $\alpha_d$, $\alpha_s$. As we will describe in Sec 2.2.1, we impose priors on some of these and constrain others. We refer to these parameters as `secondary' parameters, to distinguish them from `primary' cosmological parameters describing the primordial fluctuations.  In part of the analysis we will estimate parameters from the \arone\ spectrum alone. In this case $i=j$ and $\nu_i=\nu_0$. The model in Eq.~\ref{eqn:spectra_th} then simplifies to
\ba
\label{eq:model2}
{\cal B}_\ell^{\rm{th}} &=& {\cal B}_\ell^{\rm{CMB}} 
+ A_{\rm tSZ}{\cal B}_{0,\ell}^{\rm{tSZ}}
+ A_{\rm kSZ}{\cal B}_{0,\ell}^{\rm{kSZ}}
\nonumber\\
&& \qquad + (A_s+A_d)\left(\frac{\ell}{3000}\right)^2+ A_c {\cal B}_{0,\ell}^{\rm clust}.
\ea
This can be further simplified  to
\be
\label{eq:model3}
{\cal B}_\ell^{\rm{th}} = {\cal B}_\ell^{\rm{CMB}} 
+ A_{\rm SZ}{\cal B}_{0,\ell}^{\rm{SZ}}
+ A_p\left(\frac{\ell}{3000}\right)^2
+ A_c {\cal B}_{0,\ell}^{\rm clust},
\ee
where $A_p=A_s+A_d$ is the total Poisson power at $\ell=3000$, and $A_{\rm SZ}=A_{\rm kSZ}=A_{\rm tSZ}$ measures the total SZ power, ${\cal B}_{0,\ell}^{\rm{SZ}}={\cal B}_{0,\ell}^{\rm{tSZ}}+{\cal B}_{0,\ell}^{\rm{kSZ}}$. This is the same parameterization considered in \citet{fowler/etal:prep}, and has just three secondary parameters: $A_{\rm SZ}$, $A_p$ and $A_c$.

Using this model, we can compute the theoretical spectra for a given set of secondary parameters $\Theta$, and for a given theoretical CMB temperature power spectrum. The data power spectra are not measured at every multipole, so bandpower theoretical spectra are computed using $C_b^{\rm th,ij} =
w^{ij}_{b\ell}C_\ell^{\rm th,ij}$, where $w^{ij}_{b\ell}$ is the bandpower window function in band $b$ for cross-spectrum $ij$, described in \citet{das/etal:prep}.\footnote{Here we use the notation $w_{b\ell}$ for the bandpower window functions; \citet{das/etal:prep} uses $B_{b\ell}$.} The data power spectra have calibration uncertainties (to be described in Sec \ref{subsubsec:calbeamerr}). To account for these uncertainties we include two calibration parameters $y(\nu_i)$, for each map $i$, that scale the estimated data power spectra, ${\hat C}^{\rm ij}_b$, and their uncertainties, as
\be
C^{\rm ij}_b = y(\nu_i)y(\nu_j){\hat C}^{\rm ij}_b.
\label{eqn:cal}
\ee

The likelihood of the calibrated data is then given by 
\be
 -2 \ln\mathscr{L} = (C_b^{\rm th} - C_b)^{\rm T} 
 \mat{\Sigma}^{-1}(C_b^{\rm th} -
  C_b) + \ln \det\mat{\Sigma},
\label{eqn:like}
\ee
assuming Gaussian uncertainties on the measured bandpowers with covariance matrix $\mat{\Sigma}$.  For the 148+218 likelihood the model and data vectors $C_b^{\rm th}$ and $C_b$ contain three spectra, $C_b = [C_b^{148,148},C_b^{148,218},C_b^{218,218}]$. For the 148-only likelihood, $C_b = C_b^{148,148}$.
We use the data between  $500<\ell<10000$ for the $148\times148$ GHz auto-spectrum, but restrict the range to $1500<\ell<10000$ for the $218 \times 218$~GHz and the $148 \times 218$~GHz spectra. This range is chosen since for \arone\ at scales larger than $\ell=500$ the signal cannot be accurately separated from atmospheric noise, and for \artwo\ the maps do not converge below $\ell=1500$, described in \citet{das/etal:prep}. The bandpower covariance matrix $\mat{\Sigma}$ is described in the Appendix of \citet{das/etal:prep}, and includes correlations between the three spectra. \citet{das/etal:prep} demonstrates with Monte-Carlo simulations that the covariance is well modeled by a Gaussian distribution with negligible correlation between bands.

\subsubsection{Calibration and beam uncertainty}
\label{subsubsec:calbeamerr}

The \act\ calibration is described in \citet{hajian/etal:prep}. The \arone\ maps are calibrated using \wmap, resulting in a 2\% map calibration error in temperature units, at effective $\ell=700$. The \artwo\ maps are calibrated using observations of Uranus, with a 7\% calibration error at $\ell=1500$. The two calibration errors have negligible covariance, and are treated as independent errors. For analyses using \arone\ data alone 
we marginalize over the calibration uncertainty analytically following
\citet{ganga/ratra/sugiyama:1996} and \citet{bridle/etal:2002}. For joint analyses with the \arone\ and \artwo\ data we explicitly sample the calibration parameters $y(\nu_i)$ with 
Gaussian priors of $y(148)=1.00\pm0.02$ and $y(218)=1.00\pm0.07$. We check that 
the analytic and numerical marginalization methods give the same results 
for \arone.

The beam window functions are described in \citet{das/etal:prep}, and
are estimated using maps of Saturn. The maps are made with an 
independent pipeline to the initial ACT beam estimates 
made in \citet{hincks/etal:prep}, but produce 
consistent results. The uncertainties on the beam window functions are of order 
0.7\% for the \arone\ band and 1.5\% at \artwo. Uncertainties in the 
measured beams are incorporated using a likelihood approximation described in 
Appendix A; the magnitude of the derived uncertainties is 
consistent with \citet{hincks/etal:prep} and the uncertainties used in the parameter estimation in \citet{fowler/etal:prep}.

\subsubsection{SZ templates}
\label{subsubsec:sz_temp}

The thermal SZ template ${\cal B}^{\rm tSZ}_{0,\ell}$ describes the power from tSZ temperature fluctuations from all clusters, normalized for a universe with amplitude of matter fluctuations $\sigma_8=0.8$. There is uncertainty in the expected shape and amplitude of this signal, arising due to incomplete knowledge of the detailed gas physics that affects the integrated pressure of the clusters.  Previous cosmological studies, e.g., the ACBAR and CBI experiments, have used templates derived from hydrodynamical simulations \citep{bond/etal:2005}. The analysis for \wmap\ used the analytic Komatsu-Seljak (K-S) spectrum derived from a halo model \citep{komatsu/seljak:2001}. Recent studies for SPT have considered simulations and analytic templates from \citet{sehgal/etal:2010a} and \citet{shaw/etal:2009}. 
 
In this analysis we consider four thermal SZ templates, from \citet{sehgal/etal:2010a}, \citet*{trac/bode/ostriker:prep}, \citet{battaglia/etal:prep}, and \citet{shaw/etal:prep}. \citet*{trac/bode/ostriker:prep} constructed several templates by processing a dark matter simulation to include gas in dark matter halos and in the filamentary intergalactic medium. Their `standard' model, which was first described in \citet{sehgal/etal:2010a}, is referred to here as `TBO-1'.  It is based on the gas model in \citet*{bode/ostriker/vikhlinin:2009}, with the hot gas modeled with a polytropic equation of state and in hydrostatic equilibrium,  with star formation and feedback calibrated against observations of local clusters. This is the template considered in the ACT analysis by \citet{fowler/etal:prep}, and has a similar amplitude and shape to the K-S spectrum.

Recent state-of-the-art simulations described in \citet{battaglia/etal:prep} have been used to predict the SZ spectrum (referred to as `Battaglia'). Full hydrodynamical SPH simulations were made of large scale cosmic structure, including radiative cooling, star formation, feedback from AGN and supernovae, and non-thermal pressure support. The predicted spectrum has 2/3 the power compared to the TBO-1 spectrum and is more consistent with SZ measurements from SPT \citep{lueker/etal:prep}. It is also compatible with predictions made that AGN heating would decrease the expected SZ power \citep{roychowdhury/ruszkowski/nath:2005}.

\begin{figure}[t]
  \centering
  \resizebox{.5\textwidth}{!}{
  \hskip -0.8cm
  \plotone{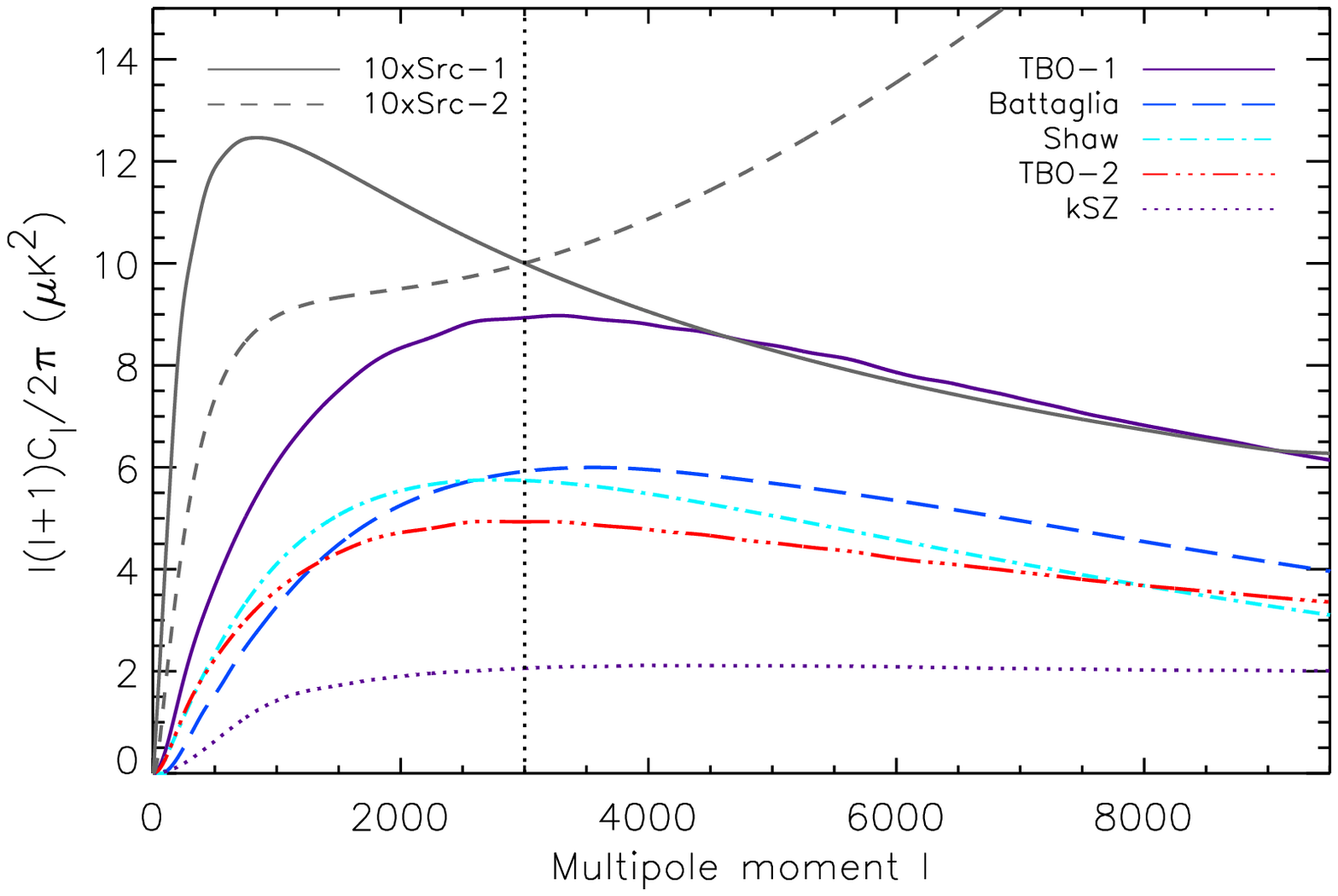} 
} 
\caption{Thermal SZ templates for four different models considered in this analysis, and a single kSZ template, normalized at \arone\ for cosmologies with $\sigma_8=0.8$. The `TBO-1' template is from \citet{sehgal/etal:2010a}, described further in \citet*{trac/bode/ostriker:prep} together with `TBO-2', derived from N-body simulations.  The `Battaglia' model is derived from hydrodynamic SPH simulations \citep{battaglia/etal:prep}. The `Shaw' model is based on an analytic halo model \citep{shaw/etal:prep}. The `kSZ' template is the kinetic SZ template in \citet{sehgal/etal:2010a}. Two clustered IR source templates considered (`Src-1' and `Src-2') are described in Sec 2.1.3 and normalized at $\ell=3000$. The IR source curves are multiplied by ten for clarity.}
  \label{fig:templates}
\end{figure}

These hydrodynamical simulations pre-date the SPT observations, but there have also been recent developments in simulating and modeling the expected SZ effect in light of the SPT results in \citet{lueker/etal:prep}, and motivated by recent observations of the intra-cluster medium (see \citet*{trac/bode/ostriker:prep} for a discussion). In a second model described in \citet*{trac/bode/ostriker:prep}, the nonthermal20 model referred to here as `TBO-2', there is 20\% non-thermal pressure support, with increased star formation  and reduced feedback, which has the effect of lowering the predicted SZ power. The `Shaw' model, described in \citet{shaw/etal:prep}, takes an analytic halo model approach, assuming hydrostatic equilibrium and a polytropic equation of state, with star formation, feedback from supernova and AGN, and energy transfer from dark matter to gas during mergers. It includes non-thermal pressure support. The spectra from each of these models are shown in Figure \ref{fig:templates}, with all models normalized to $\sigma_8=0.8$. We study constraints on all four templates. Apart from the TBO-1 template, all have similar amplitudes of ${\cal B}_{3000} \approx 5-6 \mu {\rm K}^2$ at $\ell=3000$, although the models have different amounts of star formation, feedback, and non-thermal support. 

The shape and amplitude of the expected kinetic SZ power spectrum is highly uncertain. We use the kSZ template described in \citet{sehgal/etal:2010a}, also shown in Figure \ref{fig:templates}. The corresponding template for the nonthermal20 model in \citet*{trac/bode/ostriker:prep} has a similar amplitude.
It is also consistent with predictions from second order perturbation theory \citep{hernandez-monteagudo/ho:2009}. In this analysis the contamination of the SZ signal by point sources is neglected, which is shown in \citet{lin/etal:2009} to be a good approximation for radio galaxies. \citet{lueker/etal:prep} show it is also a good assumption for IR sources for the current levels of sensitivity.

\subsubsection{Clustered source templates}
\label{subsubsec:cluster_temp}

The shape and amplitude of the power spectrum of clustered dusty galaxies are not yet well characterized \citep{knox/etal:2001,fernandez-conde/etal:2008,viero/etal:2009,hall/etal:2010}, although there have been measurements made by \citet{viero/etal:2009} from the BLAST experiment. In \citet{fowler/etal:prep} we adopted a simple power law model, with ${\cal B}_\ell\propto \ell$. In this analysis we move beyond this simple parameterization to consider two alternative model templates. The first, `Src-1', is obtained from the infrared source model described in Section 2.5.2 of \citet{sehgal/etal:2010a}. This model assumes that the IR emission traces star formation in halos at $z<3$, and that the number of IR galaxies in a given halo is proportional to its mass. For the spectral energy distribution (SED) of the galaxies, the model uses an effective graybody law in which the dust temperature is a function of the CMB temperature, but its value at $z=0$ is a free parameter. The dust emissivity spectral index and the typical IR luminosity and characteristic masses of the halos hosting IR galaxies are free parameters of the model.  
The model includes only the two-halo power spectrum given in Eq.~24 of \citet{sehgal/etal:2010a}, with contributions from pairs of galaxies in different halos. The parameters of the model have been updated from \citet{sehgal/etal:2010a} to better fit the observed BLAST power spectra at 250--500$\mu {\rm m}$.\footnote{The updated parameters are $\beta = 1.4$, $T_0 = 25.5$, $M_1 = 4 \times 10^{11}$, $M_2 = 2.5 \times 10^{12}$, $M_{\rm cool} = 5\times 10^{14}$, $L_{\star} = 2.3 \times 10^{11}$, with definitions in \citet{sehgal/etal:2010a}.} This updated template is shown in Figure \ref{fig:templates}, normalized to unity at $\ell=3000$. The shape is similar to the clustered model used in the SPT analysis by \citet{hall/etal:2010}, peaking at $\ell\simeq1000$.

We also consider the effect on our results of using an alternative template, `Src-2', that has both one-halo and two-halo power, following a halo model prescription similar to \citet{viero/etal:2009}. Dark matter halos are populated using galaxy source counts from the source model presented in \citet*{lagache/dole/puget:2003}, and halo occupation distribution parameters are tuned to fit the BLAST power spectra. This normalized template is also shown in Figure \ref{fig:templates}. At large scales this has a similar shape to the Src-1 template, but at small scales tends approximately to the ${\cal B}_\ell \propto \ell$ scaling adopted in \citet{fowler/etal:prep}, which was motivated by observations of galaxy clustering at small angles with typical correlation function $C(\theta)\propto
\theta^{-0.8}$ \citep{peebles:1980}. The two templates differ at $\ell>3000$, but at these angular scales the Poisson power is expected to dominate over the clustering term.

\subsubsection{Likelihood prescription}
\label{subsubsec:recipe}

To summarize the methods, an analysis with the `148+218' likelihood follows these steps to return the ACT likelihood for a given model:
\begin{itemize}
\item Select primary cosmological parameters, and compute a theoretical lensed CMB power spectrum ${\cal B}_\ell^{\rm{CMB}}$ using the CAMB numerical Boltzmann code \citep{lewis/challinor/lasenby:2000}. 
\item Select values for secondary parameters $\Theta=\{A_{\rm tSZ}$, $A_{\rm kSZ}$, $A_d$, $A_c$, $A_s$, $\alpha_d$, $\alpha_s\}$, and compute the total theoretical power spectra ${\cal B}_\ell^{\rm{th,ij}}$ for $148\times148$, $148\times218$ and $218\times218$ using Eq.~\ref{eqn:spectra_th}.
\item Compute the bandpower theoretical power spectra $C_b^{\rm th,ij} =
w^{ij}_{b\ell}C_\ell^{\rm th,ij}$.
\item Select values for the calibration factors for \arone\ and \artwo, and compute the likelihood using Eq.~\ref{eqn:like} for $500<\ell<10000$ for $148\times148$ and $1500<\ell<10000$ for $148\times218$ and $218\times218$.
\item Add the likelihood term due to beam uncertainty, described in Appendix A.
\end{itemize}

A large part of our analysis uses only the \arone\ spectrum. An analysis done with this `148-only' likelihood follows these steps:
\begin{itemize}
\item Select primary cosmological parameters, and compute a theoretical lensed CMB power spectrum ${\cal B}_\ell^{\rm{CMB}}$ using CAMB. 
\item Select values for secondary parameters $\Theta=\{A_{\rm SZ}$, $A_p$, $A_c\}$ and compute the total theoretical power spectrum ${\cal B}_\ell^{\rm{th}}$ at \arone\ using Eq.~\ref{eq:model}.
\item Compute the bandpower theoretical power spectrum $C_b^{\rm th} =
w_{b\ell}C_\ell^{\rm th}$.
\item Compute the likelihood using Eq.~\ref{eqn:like} for $500<\ell<10000$ for \arone.
\item Add the likelihood term due to beam uncertainty, described in Appendix A, and analytically marginalize over the calibration uncertainty.
\end{itemize}

\subsection{Parameter estimation methods}
\label{subsec:param_methods}

We use the ACT likelihood for two separate parameter investigations. The first uses the 148+218 likelihood to constrain the secondary parameters,
as our initial goal is to characterize the small-scale behavior, and investigate whether this simple model sufficiently describes the observed emission. The second uses the 148-only likelihood to constrain primary and secondary parameters. 

\subsubsection{Secondary parameters from 148 and \artwo}
\label{subsubsec:highell}

For most of the investigation with the 148+218 likelihood we fix the primary cosmological parameters to the best-fit \LCDM\ parameters estimated from \wmap, as our goal is to characterize the small-scale power observed by ACT, and check the goodness of fit of this simple model.  To map out the probability distribution for these parameters we use an MCMC method. This uses the Metropolis algorithm to sample parameters \citep{metropolis/etal:1953}, following the methodology described in \citet{dunkley/etal:2005}. 

There are seven possible secondary parameters, but we do not allow them all to vary freely. The radio sources detected at \arone, described in \citet{marriage/etal:2010a}, are observed to have typical spectral index $S(\nu)\propto \nu^{-0.5}$ in flux units. By fitting a scaled source model from \citet{toffolatti/etal:1998} to the detected sources, and using it to extrapolate to fainter sources, \citet{marriage/etal:2010a} predict a residual radio source power of $C_\ell= 2.8\pm0.3$~n${\rm K}^2$. 
Converting units, we use these measurements to impose a Gaussian prior of $A_s=4.0\pm0.4$~$\mu {\rm K}^2$, and we fix $\alpha_s=-0.5$. We also fix $A_{\rm kSZ}=A_{\rm tSZ}$, as the kSZ component is subdominant at \arone\ and the SZ models predict them to be the same for a given cosmology. The other parameters ($A_{\rm tSZ}$, $A_d$, $A_c$, and $\alpha_d$) have uniform priors with positivity imposed on the amplitudes. Parameter results are quoted using the means and 68\% confidence limits of the marginalized distributions, with 95\% upper or lower limits given when 
the distribution is one-tailed. We also quote derived parameters to indicate the power in different components at \arone\ and \artwo, for example the total power in SZ at $\ell=3000$, ${\cal B}_{3000}^{\rm SZ} \equiv ({\cal B}^{\rm kSZ}+{\cal B}^{\rm tSZ})_{3000}$.

\begin{figure*}[htb]
  \centering
  \resizebox{0.9\textwidth}{!}{
  \plotone{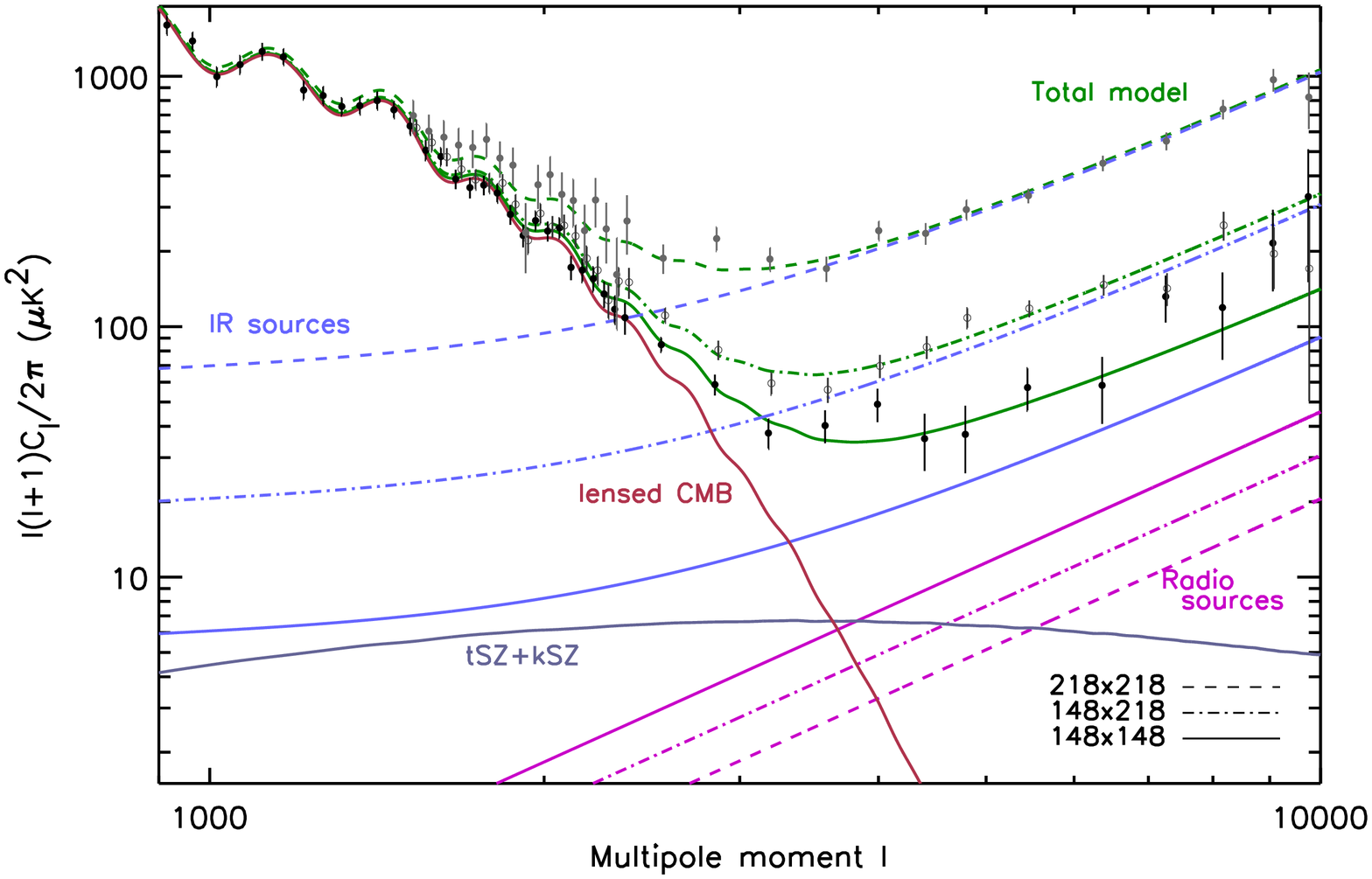} 
} 
\caption{The angular power spectrum measured by \act\ at \arone\ and \artwo\ \citep{das/etal:prep}, with the theoretical  model for CMB, SZ, and point sources best-fit to the three spectra. The lensed CMB corresponds to the \LCDM\ model with parameters derived from \wmap\ \citep{komatsu/etal:prep}. It dominates at large scales, but falls exponentially due to Silk damping. The majority of power at $\ell>3000$ comes from extragalactic point sources below a $\approx$20 mJy flux cut after masking. The radio sources  are sub-dominant, and are constrained by a source model fit to detected sources at \arone\ \citep{marriage/etal:2010a}. The infrared source emission, assumed to follow a power law, is dominated by Poisson power at small scale, but about 1/3 of the IR power at $\ell=3000$ is attributed to clustered source emission, assuming a template described in the text. The best-fit SZ (thermal and kinetic) contribution at \arone\ (assuming the TBO-1 template, \citet{sehgal/etal:2010a}) is $7 \mu {\rm K}^2$ at $\ell=3000$; the subdominant kinetic SZ also contributes at \artwo. The data spectra and errors have been scaled by best-fit calibration factors of 
$1.02^2$, $1.02\times1.09$ and $1.09^2$ for the $148\times148$, $148\times218$, and $218\times218$ spectra respectively.}
  \label{fig:l2_spectrum}
\end{figure*}

\subsubsection{Parameters from \arone}
\label{subsubsec:148_only}

In order to explore the probability distributions for a set of cosmological models, we use the 148-only likelihood for parameter estimation. The focus is on using the $1000<\ell<3000$ spectrum to improve constraints on primary cosmological parameters.  It is important that the SZ and foregound contribution be marginalized over, but we exclude the more contaminated \artwo\ data given the current uncertainties in the foreground model. To map out the distribution for cosmological parameters we use MCMC methods to 
explore the probability distributions for various cosmological models. 

We parameterize cosmological models using 
\be
\{ \Omega_b h^2, \Omega_c h^2, \Omega_\Lambda,  \Delta_{\cal R}^2,n_s,\tau \}.
\ee
These are the basic $\Lambda$CDM parameters, describing a flat universe with
baryon density $\Omega_b h^2$, cold dark matter (CDM) density $\Omega_c h^2$, 
and a cosmological constant $\Omega_\Lambda$. Primordial
perturbations are assumed to be scalar, adiabatic, and Gaussian, described by 
a power-law with spectral tilt $n_s$, and amplitude $\Delta_{\cal R}^2$, defined at pivot scale $k_0=0.002/$Mpc. We assume `instantaneous' reionization, where the universe transitions from neutral to ionized over a redshift range $\Delta z = 0.5$, with optical depth $\tau$. Reionization likely takes place more slowly \citep[e.g.,][]{gnedin:2000,trac/cen/loeb:2008}, but current CMB measurements are insensitive to this choice \citep{larson/etal:prep}.  
We also consider an additional set of primary parameters
\be
\{dn_s/d\ln k,r, N_{\rm eff}, Y_P, G\mu \},
\ee 
that describe primordial perturbations with a running scalar spectral index 
$dn_s/d \ln k$, a tensor contribution with 
tensor-to-scalar ratio $r$, a varying number of relativistic species 
$N_{\rm eff}$, varying 
primordial Helium fraction $Y_P$, and cosmic strings with tension $G\mu$, using the Nambu string template described in \citet{battye/moss:prep}. These parameters are added individually to the $\Lambda$CDM model in order to look for possible deviations from the concordance cosmology. Apart from $G\mu$ these parameters all take uniform priors, with  
positivity priors on $r$, \neff, and $Y_P$. The tensor spectral index is fixed at $n_t=-r/8$, and both the index and ratio are defined as in e.g., \citet{komatsu/etal:2009}. The CMB power spectrum from cosmic strings is 
expected to scale as $(G\mu)^2$, so we follow \citet{sievers/etal:2009} and \citet{battye/moss:prep} by parameterizing the string power using $q_{\rm str} \propto (G\mu)^2$. Limits on $G\mu$ are then derived from $q_{\rm str}$. 

We generate the lensed theoretical CMB spectra using CAMB\footnote{Version Feb 2010, with Recfast 1.5.}, and for computational efficiency set the CMB to zero above $\ell=4000$ where the contribution is subdominant, less than 5\% of the total power. 
To use the 148-only \act\ likelihood there are three secondary parameters, $A_{\rm SZ}, A_p$, and $A_c$. For this part of the analysis we use the TBO-1 and Src-1 SZ and clustered source templates, checking the effect on the primary parameters of substituting alternative templates. We also impose positivity priors on these parameters. 
We do not use any information explicitly from the \artwo\ spectrum in 
this part of the analysis, using just the 148-only likelihood, although 
results are checked using the 148+218 likelihood.
The ACT likelihood is combined with the seven-year \wmap\ data and 
other cosmological data sets. We use the 
MCMC code and methodology described in Appendix C of 
\citet{dunkley/etal:2009}, with the convergence test 
described in \citet{dunkley/etal:2005}.  A subset of results are
cross-checked against the publicly available CosmoMC code. 

To place constraints on cosmological parameters we use the 7-year \wmap\ data in combination with \act, using the \wmap\ likelihood package v4.1 described in \citet{larson/etal:prep}. \wmap\ measures the CMB over the full sky to 0.2\degree\ scales. All \wmap-only results shown for comparison use MCMC chains from LAMBDA\footnote{\url{http://lambda.gsfc.nasa.gov/}}, described in \citet{larson/etal:prep}. We follow the methodology described in \citet{komatsu/etal:prep} to consider the addition of distance measurements from astrophysical observations, on the angular diameter distances measured from Baryon Acoustic Oscillations (BAO) at $z=0.2$ and $0.35$, and on the 
Hubble constant. The Gaussian priors on the distance ratios, $r_s/D_V(z=0.2)=0.1905\pm 0.0061$ and $r_s/D_V(z=0.35)=0.1097\pm 0.0036$, are derived from 
measurements from the Two-Degree Field Galaxy Redshift Survey (2dFGRS) and the Sloan Digital Sky Survey Data
Release 7 (SDSS DR7), using a combined analysis of the two data-sets by \citet{percival/etal:2010}. The parameter 
$r_s$ is the comoving sound horizon size at the 
baryon drag epoch, and $ D_V(z) \equiv [(1+z)^2D_A^2(z)cz/H(z)]^{1/3}$ is the effective distance measure for angular diameter distance $D_A$, and Hubble parameter $H(z)$. The inverse covariance matrix is given by Eq.~5 of \citet{percival/etal:2010}. The Gaussian prior on the Hubble constant, 
$H_0=74.2\pm 3.6~{\rm km~s^{-1}~Mpc^{-1}}$, comes from 
the magnitude-redshift relation from HST observations of 
240 low-$z$ Type Ia supernovae at $z<0.1$ by \citet{riess/etal:2009}. The 
error includes both statistical
and systematic errors.

\section{High-ell SZ and Point Source Model}
\label{sec:sz_ptsrc}

\begin{table}[t]
  \centering
  \caption{ 
    Parameters describing SZ and extragalactic
  source model at \arone\ and \artwo}
  \begin{tabular}{l|cc}
\hline
Parameter\tablenotemark{a}&  148 + 218~GHz &\arone\ -only \\
\hline 
\hline
$A_{\rm tSZ}$\tablenotemark{b} 
&\ensuremath{0.62\pm 0.26} 
&$<0.77$ (95\% {\rm CL})\\%
$A_{d}$ ($\mu {\rm K}^2$)
& \ensuremath{7.8\pm 0.7} 
&$12.0\pm1.9$\\
$A_{c}$  ($\mu {\rm K}^2$)
& \ensuremath{4.6\pm 0.9} 
&$<7.4$ (95\% {\rm CL})\\
$A_{s}$ ($\mu {\rm K}^2$)\tablenotemark{c} 
& \ensuremath{4.1\pm 0.4} 
&$4.0\pm0.4$\\
$\alpha_d$\tablenotemark{d} 
& \ensuremath{3.69\pm 0.14} 
& $-$\\
\hline
$\chi^2$/dof &78/106& 29/46\\
\hline
     \end{tabular}
\tablenotetext{a} {The kSZ and tSZ coefficients are set equal, $A_{\rm kSZ}=A_{\rm tSZ}$. $A_d$, $A_c$ and $A_s$ are the ${\cal B}_{3000}$ power for Poisson infrared galaxies, clustered infrared galaxies, and Poisson radio galaxies at \arone\ respectively. The \LCDM\ parameters are not varied here.} 
\tablenotetext{b} {For the TBO-1 template. See Table 2 for other templates and conversion to SZ power.}
\tablenotetext{c} {A Gaussian prior $A_s=4.0\pm0.4$ is imposed, and index $\alpha_s=-0.5$ assumed.}
\tablenotetext{d} {The \arone-only data cannot constrain the IR point source index $\alpha_d$.}
\label{table:highell_params}
\end{table}

In this section we determine the goodness of fit of the SZ and point source model described in Section \ref{subsec:act_like} to the \act\ \arone\ and \artwo\ power spectra, and estimate its parameters. This uses the 148+218 likelihood summarized in Sec \ref{subsubsec:recipe}, initially holding the \LCDM\ model fixed to the primary CMB with parameters given in \citet{komatsu/etal:prep}.
The best-fit model is a good fit to the three \act\ power spectra over the full angular range $500<\ell<10000$ ($\chi^2=78$ for 106 degrees of freedom), with  constraints on parameters given in Table \ref{table:highell_params} for the TBO-1 SZ template and Src-1 source template. The spectra are shown in Figure \ref{fig:l2_spectrum}, with the estimated components indicated at each frequency. The mean calibration factors, defined in Eq.~\ref{eqn:cal}, are $1.02$ and $ 1.09$ for \arone\ and \artwo\ respectively. These are consistent with the expected values at the 1-1.2$\sigma$ level. The best-fitting 1.09 factor is driven by the $\ell<2500$ part of the $148\times218$ cross-spectrum, where the primary CMB dominates. At $\ell=3000$, about half the power at \arone\ is from the primary CMB (27 out of $50~\mu {\rm K}^2$), with the remainder divided among SZ, IR Poisson and clustered power, and radio Poisson power (4-8 $\mu {\rm K}^2$ in each component). At \artwo, only about 15\% of the power comes from the primary CMB at $\ell=3000$ (27 out of 170 $\mu {\rm K}^2$). Half of the power is attributed to Poisson IR sources, the remaining approximately 35\% to power from clustered IR sources. The model fits the cross-spectrum, indicating that a similar population of galaxies is contributing at both frequencies. 

\subsection{Constraints on SZ power}
\label{subsec:sz_power}

\begin{table}[t]
  \centering
  \caption{ 
    Constraints on SZ emission}
  \begin{tabular}{l|cccc}
\hline
Template\tablenotemark{a}   & $A_{\rm tSZ}$\tablenotemark{b}    & ${\cal B}^{\rm SZ}_{3000}$\tablenotemark{c} & $\sigma_8^{\rm SZ,7}$  &$\sigma_8^{\rm SZ,9}$ \\
&  &  ($\mu {\rm K}^2$) & $0.8\times(A_{\rm tSZ}^{1/7})$ & $0.8\times(A_{\rm tSZ}^{1/9})$\\
\hline 
\hline
TBO-1
&$0.62\pm0.26$
&$6.8\pm2.9$
&$0.74\pm0.05$
&$0.75\pm0.04$
\\
TBO-2
&$0.96\pm0.43$
&$6.7\pm3.0$
&$0.78\pm0.05$
&$0.79\pm0.04$
\\
Battaglia
&$0.85\pm0.36$
&$6.8\pm2.9$
&$0.77\pm0.05$
&$0.78\pm0.04$
\\
Shaw
&$0.87\pm0.39$
&$6.8\pm3.0$
&$0.77\pm0.05$
&$0.78\pm0.04$
\\
\hline
\end{tabular}
\tablenotetext{a}{Templates are from \citet{sehgal/etal:2010a}, \citet*{trac/bode/ostriker:prep}, \citet{battaglia/etal:prep}, and \citet{shaw/etal:prep}.}
\tablenotetext{b}{We required $A_{\rm kSZ}=A_{\rm tSZ}$, as defined in Eqs.~6-7.}
\tablenotetext{c}{Total tSZ and kSZ power at \arone, as defined in Eq.~8.}
\label{table:sz_param}
\end{table}

\label{subsec:sz_limit}
\begin{figure}[htb]
  \centering
  \resizebox{0.5\textwidth}{!}{
  \plotone{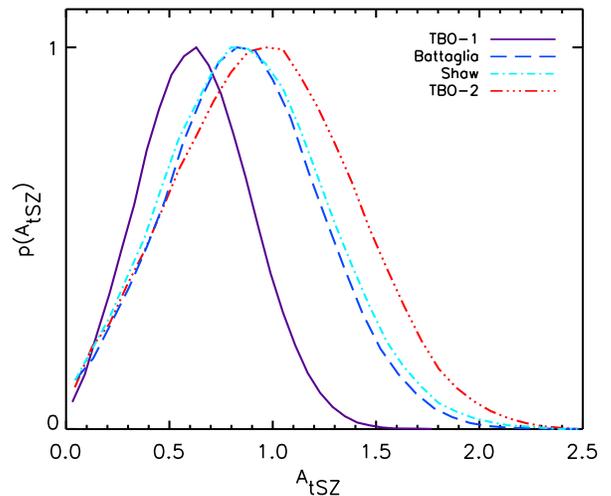} 
} 
\caption{One-dimensional marginalized distributions for the estimated thermal SZ power in the \act\ power spectra. There is evidence at the 95\% CL level for non-zero SZ power. The value $A_{\rm tSZ}=1$ corresponds to the predicted thermal SZ amplitude in a universe with $\sigma_8=0.8$. The four curves correspond to the four SZ templates shown in Figure 1; the  TBO-1 template results in a lower value, although all are consistent with $A_{\rm tSZ}=1$ at the 95\% CL. The total SZ power (including kSZ) at \arone\ and $\ell=3000$ for all the templates is consistent, with $\ell(\ell+1)C^{\rm SZ}_\ell/2\pi=7\pm3$~$\mu {\rm K}^2$.} 
  \label{fig:sz_param}
\end{figure}

\begin{figure*}[htb]
  \centering
  \resizebox{1.\textwidth}{!}{
  \plotone{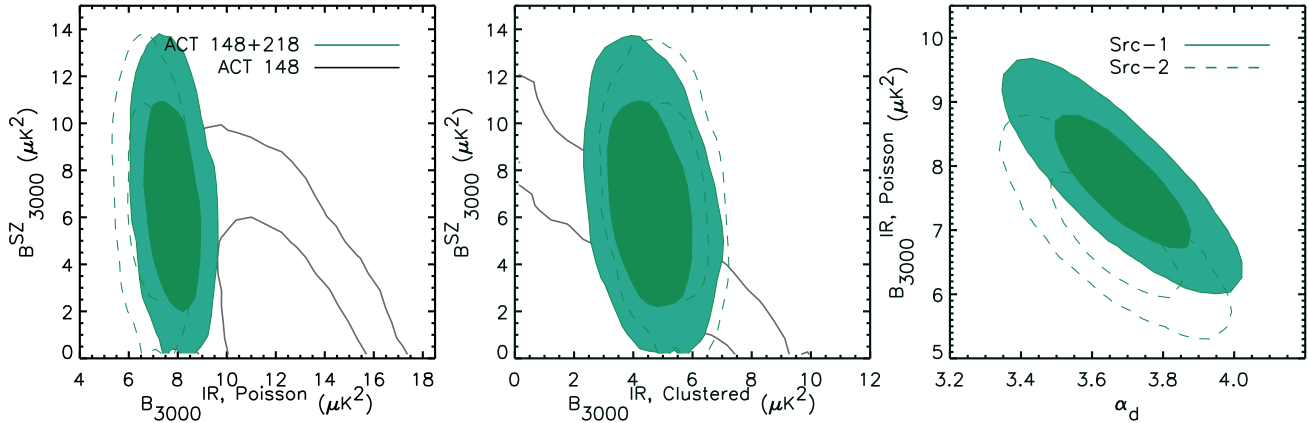} 
}
\caption{Marginalized distributions (68\% and 95\% CL) for parameters describing the SZ and point source emission in the \act\ power spectra. {\it Left and center}: The degeneracies between the total SZ power, ${\cal B}_{\ell}^{\rm SZ}\equiv \ell(\ell+1)C_\ell^{\rm SZ}$, and the infrared point source power,  ${\cal B}_{\ell}^{\rm IR}$, at \arone\ and $\ell=3000$ (solid unfilled contours), are broken with the addition of \artwo\ data (solid filled contours). Both the Poisson and clustered IR power are shown, for two different clustered source templates (solid and dashed contours). A clustered source component is required to fit the multi-frequency data at 5$\sigma$ significance. {\it Right}: The Poisson dust power and the index $\alpha_d=3.69\pm0.14$ (power law in flux between \arone\ and \artwo, and unconstrained from \arone\ alone) are anti-correlated; the index indicates a dust emissivity of $\beta\approx 1.7$.}
  \label{fig:5d_params}
\end{figure*}

Using the multi-frequency spectra, power from SZ fluctuations is detected at more than 95\% CL, with estimated $A_{\rm tSZ}$ for each template (TBO-1, TBO-2, Battaglia, and Shaw) given in Table \ref{table:sz_param} and shown in Figure \ref{fig:sz_param}, marginalized over point source parameters. The estimated SZ power at $\ell=3000$ is robust to varying the SZ template, with total SZ power (tSZ plus kSZ) estimated to be
\be
{\cal B}^{\rm SZ}_{3000}=6.8\pm2.9~\mu {\rm K}^2.
\ee
The estimated template amplitude, $A_{\rm tSZ}$, varies from $0.62\pm0.26$ for the TBO-1 template, to $0.96\pm0.43$ for the TBO-2 template. Note that $A_{\rm kSZ}$ is fixed equal to $A_{\rm tSZ}$ in these cases, with amplitudes defined in Eqs.~6-7. For the TBO-1 template, the mean amplitude is lower than expected for a universe with $\sigma_8=0.8$ ($A_{\rm tSZ}=1$), but not significantly. This is consistent with observations by SPT \citep{lueker/etal:prep}, and is an improvement over the initial estimate of $A_{\rm tSZ}<1.6$ at 95\% CL from the ACT power spectrum presented in \citet{fowler/etal:prep}. Assuming that $\sigma_8$ is within the limits estimated from primary CMB data, e.g. from \citet{komatsu/etal:prep}, the amplitude is somewhat more consistent for the TBO-2, Battaglia, and Shaw templates, that include more detailed gas physics, with $A_{\rm tSZ}=1$ within the 68\% CL for these templates. In all these cases we have held the primary CMB parameters fixed. For a single test case we marginalize over the 6 primary \LCDM\ parameters in addition to the secondary parameters. This marginalization results in an increase in the mean value of ${\cal B}^{\rm SZ}_{3000}$ of $0.5~\mu {\rm K}^2$ (a 0.2$\sigma$ change), but a negligible increase in the uncertainty. 

The number of clusters, and therefore the expected SZ power, is a strong function of the amplitude of matter fluctuations, quantified by $\sigma_8$ \citep{komatsu/kitayama:1999}. In our model we scale the SZ templates by an overall amplitude, and would like to infer an estimate for $\sigma_8$ from $A_{\rm tSZ}$. In \citet{fowler/etal:prep} we assumed a seventh power scaling, with $A_{\rm tSZ}\propto \sigma_8^7$ \citep{komatsu/seljak:2002}, giving an upper limit of $\sigma^{\rm SZ}_8<0.84$ at 95\% CL, for $A_{\rm tSZ}<1.6$. However, the exact scaling of the shape and amplitude with cosmology, and in particular with $\sigma_8$, is model dependent and not precisely known \citep{lueker/etal:prep,battaglia/etal:prep,trac/bode/ostriker:prep}. For the TBO templates the combined tSZ and kSZ signal scales close to the 7th power, with the tSZ varying approximately as the 8th power \citep*{trac/bode/ostriker:prep}. To bound the possible range we compute two limits, assuming the tSZ part of the template varies as either $\sigma_8^7$ or $\sigma_8^9$. The estimated values for $\sigma^{\rm SZ}_8$ in these cases are given in Table \ref{table:sz_param}.

No detections have yet been made of the kinetic SZ power spectrum. From SPT observations a 95\% upper limit on ${\cal B}_{3000}^{\rm kSZ}$ of $13 \mu {\rm K}^2 $ was estimated \citep{hall/etal:2010}. If we allow the kSZ amplitude to be varied independently of the thermal SZ amplitude, we find an upper limit from the ACT data on the kinetic SZ contribution of 
\be
{\cal B}_{3000}^{\rm kSZ} <8 \mu {\rm K}^2~(95\%~{\rm CL}).
\ee
This is consistent with predictions by \citet{iliev/etal:2008} of a $2 \mu {\rm K}^2$ Ostriker-Vishniac signal at $\ell= 3000$ and a $3 \mu {\rm K}^2$ post-reionization kSZ signal, but would exclude models with higher levels of kSZ from patchy reionization. 

The estimated SZ power is a small signal, less than 10~$\mu {\rm K}^2$, and is correlated with other parameters. We therefore investigate the dependence of the constraint on the priors imposed on other parameters. The SZ power is not strongly correlated with the IR point source parameters when the \artwo\ data are included, as shown in Figure \ref{fig:5d_params}, and so using the `Src-2' clustered source template in place of `Src-1' has a negligible effect. 
There is some correlation with the radio source power, as this term also contributes predominantly at \arone. Changing the radio index to $\alpha_s=0$ has little effect, but broadening the radio prior to $A_s=4\pm2$ does reduce the significance of the SZ detection to $0.48\pm0.27$ using the TBO-1 template; the SZ is anti-correlated with the radio power, and relaxing the prior allows a larger radio component at \arone. This indicates the importance of the radio source characterization in \citet{marriage/etal:2010a} for estimating the SZ power at \arone. A modest increase in the prior to $A_s=4\pm0.8$ has a negligible effect.

If we restrict the analysis to \arone\ alone we find consistent results, 
  with ${\cal B}^{\rm SZ}_{3000}<7.8~\mu {\rm K}^2$ at 95\% CL. The best-fit model has $\chi^2=29$ for 46 degrees of freedom. In this case we cannot distinguish between the SZ and clustered source components; the joint constraint shown in Figure \ref{fig:5d_params} shows that similar $\ell=3000$ power limits are placed on both components; marginalizing over a clustered term has little effect on the estimated SZ amplitude. The mean IR Poisson term is higher in this case, as there are more models fitting the \arone\ data alone with low SZ and clustered source power. The multi-frequency information then breaks this degenerecy and more tightly constrains the Poisson power. 

\begin{figure*}[t]
  \centering
  \resizebox{1.\textwidth}{!}{
  \plotone{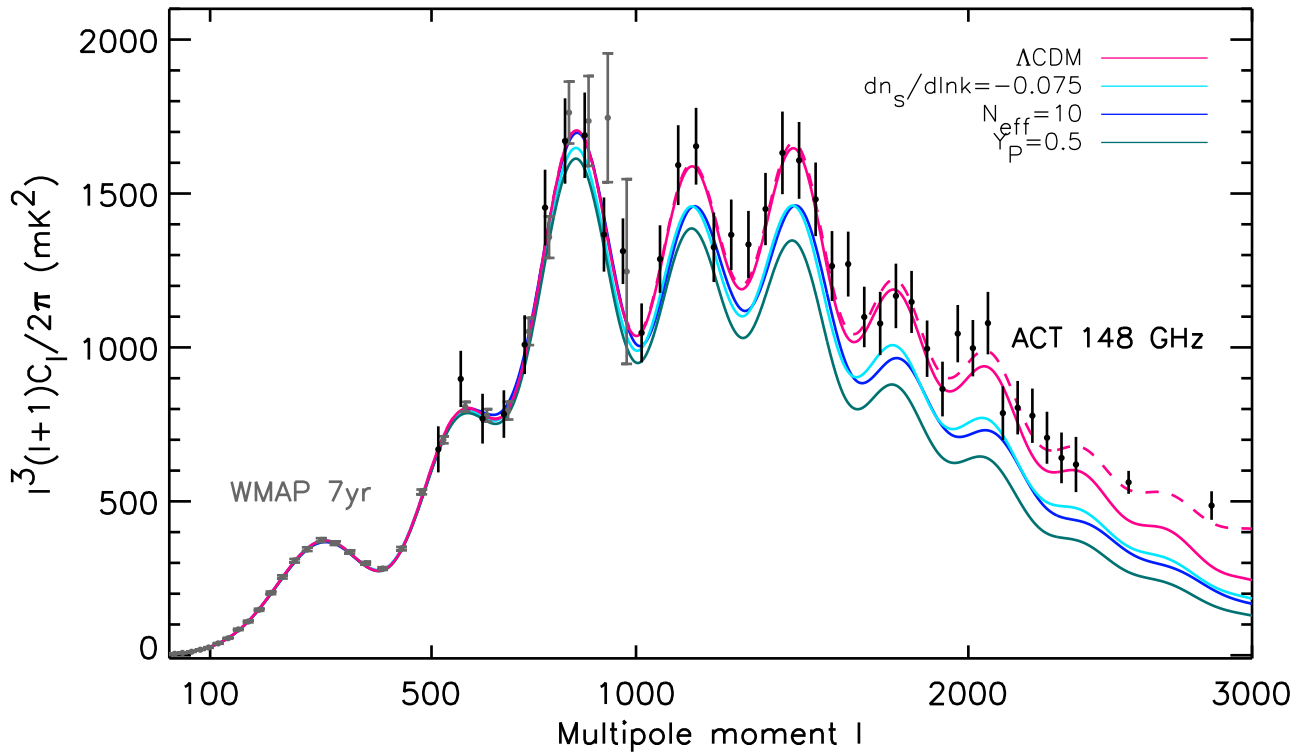} 

} 
\caption{The power spectrum measured by \act\ at \arone, scaled by $\ell^4$, over the range dominated by primordial CMB ($\ell<3000$). The spectrum is consistent with the \wmap\ power spectrum over the scales $500<\ell<1000$, and gives a measure of the third to seventh acoustic peaks. The best-fit \LCDM\ cosmological model is shown, and is a good fit to the two datasets. At $\ell>2000$ the contribution from point soures and SZ becomes significant (dashed shows the total best-fit theoretical spectrum; solid is lensed CMB).  Three additional theoretical models for the primordial CMB are shown with \neff=10 relativistic species, $^4$He fraction $Y_p=0.5$, and running of the spectral index $dn_s/d \ln k=-0.075$. They are consistent with \wmap\ but are excluded at least at the 95\% level by the \act\ data.}
  \label{fig:l4_spectrum}
\end{figure*}

\subsection{Unresolved point source emission}
\label{subsec:ptsrc}

\begin{table}[t]
  \centering
  \caption{
    Derived constraints on unresolved 
  IR source emission\tablenotemark{a}}
  \begin{tabular}{ll|cc}
\hline
&  & \arone\ & \artwo\ \\
\hline 
\hline
Poisson 
&${\cal B}_{3000}$ ($\mu {\rm K}^2$)\tablenotemark{b}
&$7.8\pm0.7\pm0.7$
&$90\pm5\pm10$
\\
&$C_\ell$(n${\rm K}^2$)
&$5.5\pm0.5\pm0.6$
&$63\pm3\pm6$
\\
&$C_\ell$ (${\rm Jy}^2\,{\rm sr}^{-1}$) 
&$0.85\pm0.08\pm0.09$
&$14.7\pm0.7\pm1.8$
\\
&&&\\
Clustered 
&${\cal B}_{3000}$  ($\mu {\rm K}^2$)\tablenotemark{c}
&$4.6\pm0.9\pm0.6$
&$54\pm12\pm5$
\\
&&&\\
Total IR
&${\cal B}_{3000}$  ($\mu {\rm K}^2$)
&$12.5\pm1.2$
&$144\pm13$
\\
\hline
     \end{tabular}
\tablenotetext{a}{The two errors indicate statistical uncertainty and a systematic error due to clustered template uncertainty.}
\tablenotetext{b}{Equivalent to the parameter $A_d$ for \arone.}
\tablenotetext{c}{Equivalent to the parameter $A_c$ for \arone.}
\label{table:dust_params}
\end{table}

The power spectrum measures fluctuations due to point sources below a 
flux cut of approximately 20~mJy. This is not an exact limit since the point source mask is constructed from sources with signal-to-noise ratio greater than 5 \citep{marriage/etal:2010a}. The point source power observed at \arone\ and \artwo\ has both synchrotron emission from radio galaxies, and IR emission from dusty galaxies. At \arone\ the point source power, after removal of 5$\sigma$ sources, is inferred to be split in ratio roughly 1:2 between radio and IR galaxies. Since we impose a prior on the residual radio power from \citet{marriage/etal:2010a}, we do not learn new information about this component from the power spectrum. At \artwo\ the point source power is dominated by IR dust emission. The IR Poisson power is estimated to be $A_d=7.8\pm0.7~\mu {\rm K}^2$, with derived Poisson IR power at \arone\ and \artwo\ given in Table \ref{table:dust_params}.  A clustered component is required to fit the data, with $A_c=4.6\pm0.9~\mu {\rm K}^2$, corresponding to power at \artwo\ of ${\cal B}^{218}_{3000}=54\pm12~\mu {\rm K}^2$. A model with no clustered component has a poorer fit to the data by $\Delta \chi^2=28$, indicating a detection of clustering at the 5$\sigma$ level. It is the \artwo\ power spectrum that provides this detection; the \arone\ spectrum is consistent with no clustered component. 

In flux units, the effective index of unresolved IR emission is 
\be
\alpha_d = 3.69\pm0.14
\ee
between \arone\ and \artwo, where $S(\nu) \propto \nu^{\alpha}$. The dust index and Poisson amplitude are anti-correlated, shown in Figure \ref{fig:5d_params}. This index estimate agrees with observations by SPT, who find $\alpha=3.9\pm0.3$ for the Poisson component, and $3.8\pm1.2$ for the clustered component  over the same frequency range \citep{hall/etal:2010}. A property that can be derived from the effective index, $\alpha$, is the dust emissivity index, $\beta$. For galaxies at redshift $z=0$ the dust emission can be described by a modified blackbody, $S(\nu) \propto \nu^\beta B_\nu(T_d)$, for dust temperature $T_d$. In the Rayleigh-Jeans (RJ) limit the flux then approximates to $S(\nu)\propto \nu^{\beta+2}T_d$, with $\beta=\alpha-2$. Using this relation gives a dust emissivity index measured by ACT of $\beta=1.7\pm0.14$, consistent with models \citep[e.g.,][]{draine:2003}. However, the RJ limit is not expected to be as good an approximation for redshifted graybodies \citep[e.g.,][]{hall/etal:2010}, adding an uncertainty to $\beta$ of up to $\simeq0.5$. This should also be considered an effective index, given the likely temperature variation within each galaxy.  

We test the dependence of these constraints on choices made in the likelihood, using the same set of tests described in Sec \ref{subsec:sz_power}. The estimated IR source parameters do not depend strongly on the SZ template chosen, with less than 0.1$\sigma$ change if we use the Battaglia or TBO-1 SZ template. If the radio source index is set to $\alpha_s=0$ instead of $-0.5$ there is a $\simeq 0.3\sigma$ reduction in the IR Poisson source power at \arone, and a $0.2\sigma$ increase in the spectral index. As found in Sec \ref{subsec:sz_power}, if the radio source power uncertainty is doubled from $A_s = 4\pm0.4~\mu {\rm K}^2$ to $4\pm0.8~\mu {\rm K}^2$  there is only a $0.1\sigma$ effect. More radio source power can be accommodated in \arone\ by increasing the width of the radio prior to $4\pm2~\mu {\rm K}^2$, resulting in a decrease in IR Poisson power at \arone\ of $\simeq 1 \sigma$, and a corresponding increase in the IR index by $\simeq0.8\sigma$, but this scenario is disfavored by the radio source counts presented in \citet{marriage/etal:2010a}. 

Substituting the alternative halo-model `Src-2' clustered source template reduces the estimated IR Poisson power by almost $1\sigma$. In this case the one-halo term contributes at small scales, transferring power from the Poisson to the clustered component. Given our uncertainty in the clustered model, we adopt this difference as an additional systematic error on the Poisson source levels, shown in Table \ref{table:dust_params}. In this simple model we have also assumed that the clustered and Poisson components trace the same populations with the same spectral index.  The goodness of fit of the simple model supports this assumption. The detected clustering levels are compatible with the detections by the BLAST experiment \citep{viero/etal:2009}, and will be explored further in future work.

\begin{figure}[t]
  \centering
\hskip -0.4cm
  \resizebox{.5\textwidth}{!}{
  \plotone{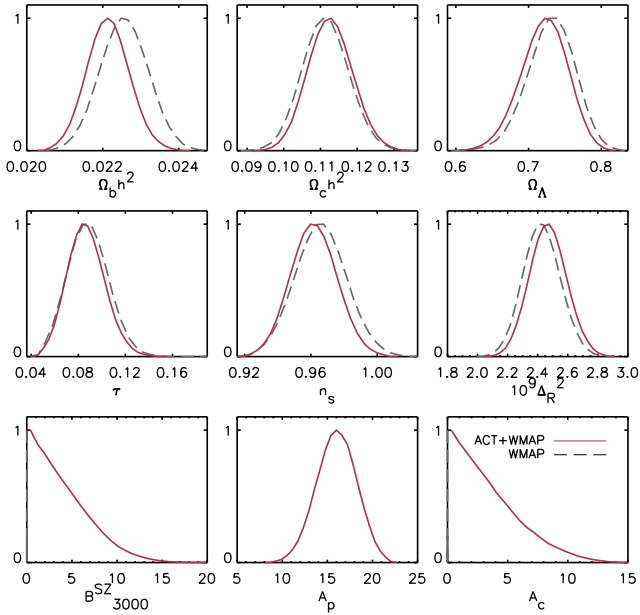} 
}
\caption{One-dimensional marginalized distributions for the 6 \LCDM\ parameters (top two rows) derived from the \act+\wmap\ combination, compared to \wmap\ alone. The bottom row shows 3 secondary parameters from the \act+\wmap\ data. With the addition of \act\ data a model with $n_s=1$ is disfavored at the 3$\sigma$ level.}
 \label{fig:lcdm_params}
\end{figure}

\section{Cosmological parameter constraints}
\label{sec:cosmo_params}

In this section we use the 148-only ACT likelihood to estimate primary cosmological parameters, in combination with \wmap\ and cosmological distance priors. Following the prescription in Section \ref{subsubsec:recipe} we marginalize over three secondary parameters to account for SZ and point source contamination. We conservatively exclude the \artwo\ data from this part of the analysis, to avoid drawing conclusions that could depend on the choice of model for the point source power.

\subsection{The \LCDM\ model}

The best-fit \LCDM\ model is shown in Figure \ref{fig:l4_spectrum}, using the combination $\ell^4C_\ell$ to highlight the acoustic peaks in the Silk damping regime.  The estimated parameters for the \act+\wmap\  combination, given in Table \ref{table:lcdm_params} and shown in Figure \ref{fig:lcdm_params}, agree to within $0.5\sigma$ with the \wmap\ best-fit. The spectral index continues to lie below the scale invariant $n_s=1$, now at the $3\sigma$ level from the CMB alone, with $n_s=\ensuremath{0.962\pm0.013}$. This supports the inflationary scenario for the generation of primordial fluctuations \citep*{mukhanov/chibisov:1981,hawking:1982,starobinsky:1982,guth/pi:1982,bardeen/steinhardt/turner:1983,mukhanov/feldman/brandenberger:1992} and is possible due to the longer lever arm from the extended angular range probed by \act. With the addition of BAO and $H_0$ data, the significance of $n_s<1$ is increased to 3.3$\sigma$, with statistics given in Table \ref{table:lcdm+all_params}.

The \LCDM\ parameters are not strongly correlated with the three secondary parameters ($A_c$, $A_p$, $A_{\rm SZ}$), as there is limited freedom within the model to adjust the small-scale spectrum while still fitting the \wmap\ data. We also find consistent results if the 148+218 ACT likelihood is used in place of the 148-only likelihood.

Evidence for the gravitational lensing of the primary CMB signal is investigated in the companion ACT power spectrum paper \citep{das/etal:prep}. A lensing parameter, $A_L$, is marginalized over that scales the lensing potential from $C_\ell^{\Psi}$ to $A_L C_\ell^{\Psi}$, as described in \citet{calabrese/etal:2008}. An unlensed CMB spectrum would have $A_L=0$, and the standard lensing case has $A_L=1$. \citet{reichardt/etal:2009} reported a detection of lensing from ACBAR; in \citet{calabrese/etal:2008} this was interpreted as a non-zero detection of the parameter $A_L$, with mean value higher than expected, $A_L=3.1^{+1.8}_{-1.5}$ at 95\% CL; \citet{reichardt/etal:2009} estimate $A_L=1.4^{+1.7}_{-1.0}$ at 95\% CL from the same ACBAR data.
With the ACT power spectrum combined with 7-year \wmap\ data, \citet{das/etal:prep} report the measure
\be
A_L=1.3^{+0.5+1.2}_{-0.5-1.0} ~(68, 95\%~ {\rm CL}),
\ee
with mean value within $1\sigma$ of the expected value. The goodness of fit of an unlensed CMB model has $\Delta \chi^2 = 8$ worse than the best-fit lensed case, indicating a 2.8$\sigma$ detection of lensing. The marginalized distribution for $A_L$ from ACT+\wmap, together with the standard lensed ($A_L=1$) and unlensed spectra ($A_L=0$),  are shown in \citet{das/etal:prep}. The measurement adds support to the standard cosmological model governing the growth rate of matter fluctuations over cosmic time, and by extracting information beyond the two-point function these measurements are expected to be improved.

\begin{table*} [t] 
\caption{\small{\LCDM\ and extended model parameters and 68\% confidence intervals from the \act\ 2008 data combined with seven-year \wmap\ data.}}
\begin{center}
\begin{tabular}{llcccccc}
\hline
\hline
&Parameter\tablenotemark{a}  & \LCDM & \LCDM & \LCDM & \LCDM & \LCDM  & \LCDM \\
& &  & + $dn_s/d\ln k$ & + $r$ & + $N_{\rm eff}$ & + $Y_P$  & + $G\mu$\\
\hline
\hline
Primary &$100\Omega_b h^2$ 
& \ensuremath{2.214\pm 0.050} 
& \ensuremath{2.167\pm 0.054} 
& \ensuremath{2.246\pm 0.057}
& \ensuremath{2.252\pm 0.055} 
& \ensuremath{2.236\pm 0.052}
& \ensuremath{2.240\pm0.053}
\\
\LCDM&$\Omega_c h^2$ 
& \ensuremath{0.1127\pm 0.0054} 
& \ensuremath{0.1214\pm 0.0074} 
& \ensuremath{0.1099\pm 0.0058}
& \ensuremath{0.152\pm 0.025} 
& \ensuremath{0.1166\pm 0.0061}
& \ensuremath{0.1115\pm0.0055}\\
&$\Omega_\Lambda$ 
& \ensuremath{0.721\pm 0.030} 
& \ensuremath{0.670\pm 0.046} 
& \ensuremath{0.738\pm 0.030}
& \ensuremath{0.720\pm 0.030} 
& \ensuremath{0.711\pm 0.031}
& \ensuremath{0.730\pm0.029}\\
&$n_s$ 
& \ensuremath{0.962\pm0.013} 
& \ensuremath{1.032\pm0.039} 
& \ensuremath{0.974\pm0.016}
& \ensuremath{0.993\pm0.021} 
& \ensuremath{0.974 \pm 0.015}
& \ensuremath{0.963\pm0.013}\\
&$\tau$ 
& \ensuremath{0.087\pm 0.014} 
& \ensuremath{0.092\pm 0.016} 
& \ensuremath{0.087\pm 0.015}
& \ensuremath{0.089\pm 0.015} 
& \ensuremath{0.087\pm 0.015}
& \ensuremath{0.087\pm0.015}\\
&$10^9\Delta_{\cal R}^2$ 
& \ensuremath{2.47\pm 0.11} 
& \ensuremath{2.44\pm 0.11} 
& \ensuremath{2.37\pm 0.13}
& \ensuremath{2.40\pm 0.12}
& \ensuremath{2.45\pm 0.11}
& \ensuremath{2.43\pm0.11}\\
\hline
Extended &$dn_s/d\ln k$ & &\ensuremath{-0.034\pm 0.018} &&& \\
&$r$ & & &\ensuremath{<0.25}&& \\
&$N_{\rm eff}$ & && &\ensuremath{5.3\pm 1.3}&\\
&$Y_P$ &&&&&\ensuremath{0.313\pm 0.044} \\
&$G\mu$ &&&&&&\ensuremath{<1.6}$\times 10^{-7}$ \\
\hline
Derived & $\sigma_8$ 
& \ensuremath{0.813\pm 0.028} 
& \ensuremath{0.841\pm 0.032} 
& \ensuremath{0.803\pm 0.030}
& \ensuremath{0.906\pm 0.059} 
& \ensuremath{0.846\pm 0.035}
& \ensuremath{0.803\pm0.029}\\
&$\Omega_m$ 
& \ensuremath{0.279\pm 0.030} 
& \ensuremath{0.330\pm 0.046} 
& \ensuremath{0.262\pm 0.030}
& \ensuremath{0.280\pm 0.030}
& \ensuremath{0.289\pm 0.031}
& \ensuremath{0.270\pm0.029}\\
&$H_0$ 
& \ensuremath{69.7\pm2.5} 
& \ensuremath{66.1\pm3.0} 
& \ensuremath{71.4\pm2.8}
& \ensuremath{78.9\pm5.9} 
& \ensuremath{69.5\pm2.3}
& \ensuremath{70.6\pm2.5}\\
\hline
Secondary&$B^{\rm SZ}_{3000}$ ~$(\mu {\rm K}^2)$
& \ensuremath{<10.2} 
& \ensuremath{<12.3} 
& \ensuremath{<10.0}
& \ensuremath{<12.1}
& \ensuremath{<13.0}
& \ensuremath{<8.8}\\
&$A_p$~$(\mu {\rm K}^2)$ 
& \ensuremath{16.0\pm 2.0} 
& \ensuremath{14.9\pm 2.2} 
& \ensuremath{16.0\pm2.0}
& \ensuremath{15.1\pm 2.1} 
& \ensuremath{15.0\pm 2.1}
& \ensuremath{16.1\pm1.9}\\
&$A_c$ ~$(\mu {\rm K}^2)$ 
& \ensuremath{<8.7} 
& \ensuremath{<10.4} 
& \ensuremath{<8.0}
& \ensuremath{<11.1}
 & \ensuremath{<11.2}
& \ensuremath{<7.4}\\
\hline
&$-2\ln {\mathscr L}$ 
&7500.0
&7498.1
&7500.1
&7498.7
&7498.8
&7500.1
\\
\hline
\hline
\end{tabular}
\tablenotetext{a}{For one-tailed distributions, the upper 95\% CL is given. For two-tailed distributions the 68\% CL are shown.}
\label{table:lcdm_params}
\end{center}
\end{table*}

\begin{figure}[t]
\hskip -0.4cm
  \resizebox{.53\textwidth}{!}{
  \plotone{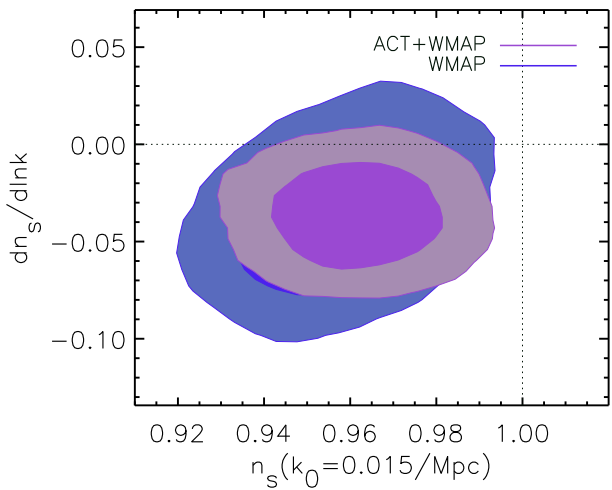} 
}
\caption{ Two-dimensional marginalized limits (68\% and 95\%) for the spectral index, $n_s$, 
 plotted at the pivot point $k_0=0.015/\rm Mpc$, and the running of the index $dn_s/d\ln k$, for \act+\wmap, compared to \wmap. This model has no tensor fluctuations. A negative running is preferred, but the data are consistent with a power-law spectral index, with $dn_s/d\ln k=0$. }
  \label{fig:running}
\end{figure}

\subsection{Inflationary parameters}
\label{subsec:infl_params}

\subsubsection{Running of the spectral index}
\label{subsec:running}

We constrain a possible deviation from power-law primordial fluctuations using the running of the index, $dn_s/d\ln k$, with curvature perturbations described by
\be
\Delta^2_{\cal R}(k) 
= \Delta^2_{\cal
R}(k_0)\left(\frac{k}{k_0}\right)^{n_s(k_0)-1+\frac12\ln(k/k_0)dn_s/d\ln
k}. 
\ee
The spectral index at scale $k$ is related to the index at pivot point $k_0$ by
\be
n_s(k)=n_s(k_0)+ \frac{dn_s}{d\ln k} \ln \left(\frac{k}{k_0}\right).
\label{eqn:running}
\ee
The simplest inflationary models predict that the running of the
spectral index with scale should be small 
\citep[see e.g.,][]{kosowsky/turner:1995,baumann/etal:2009}, 
and the detection of a scale dependence 
would provide evidence for alternative models for the early universe. 
Cosmological 
constraints on deviations from scale invariance have been considered recently by 
e.g., \citet{easther/peiris:2006,kinney/kolb/melchiorri:2006,shafieloo/souradeep:2008,verde/peiris:2008,reichardt/etal:2009}, using various parameterizations. 
With CMB data alone, the seven-year \wmap\ data show no evidence for significant running, with $dn_s/d\ln k= -0.034\pm0.026$, and $-0.041\pm0.023$  when combined with ACBAR and QUAD data \citep{komatsu/etal:prep}.
With the measurement of the power spectrum at small scales by ACT, we find
\be
\ensuremath{dn_s/d\ln{k}} = \ensuremath{-0.034\pm 0.018}~(68\%~{\rm CL})
\ee
and $dn_s/d\ln{k}$ =\ensuremath{-0.024\pm 0.015} including BAO+$H_0$. The estimated parameters are given in Tables \ref{table:lcdm_params} and \ref{table:lcdm+all_params}. Parameters are sampled using a pivot point $k_0=0.002/$Mpc for the spectral index. This choice of pivot point results in the index being strongly anti-correlated with the running, with $n_s(0.002)$ = \ensuremath{1.032\pm0.039}. In Figure \ref{fig:running} we show the index and its running at a 
decorrelated pivot point $k_0=0.015/$Mpc, chosen to minimize the correlation between the two parameters \citep*{cortes/liddle/mukherjee:2007}. The relation between the index at these two pivot points is
\ba
n_s(k_0=0.015/{\rm Mpc})&=&n_s(k_0=0.002/{\rm Mpc}) \nonumber \\
&&+  \ln(0.015/0.002)\frac{dn_s}{d\ln k},
\ea
with other cosmological parameters unchanged. The running prefers a negative value at 1.8$\sigma$, indicating enhanced damping at small scales, but there is no statistically significant deviation from a power law spectral index. 

We choose this model to investigate the sensitivity of the constraints to choices made in the likelihood, as $dn_s/d \ln k$ is more 
sensitive to the small-scale spectrum, and may be affected by the modeling of the point source and SZ contributions. We find less than $0.1\sigma$ variation in primordial parameters if we substitute alternative shapes for the SZ and clustered source templates, or limit the analysis to the $\ell<5000$ data. The beam is measured sufficently well over the angular range of interest that results are not changed if the beam uncertainty is neglected.  These tests are described further in Appendix \ref{sec:like_sensit}, and give us confidence that the errors are not dominated by systematic effects.
 
\begin{table*} [t] 
\begin{center}
\caption{\small{\LCDM\ and extended model parameters and 68\% confidence intervals from the \act\ 2008 data combined with seven-year \wmap\ data, and measurements of $H_0$ and BAO}}
\begin{tabular}{llcccc}
\hline
\hline
&Parameter\tablenotemark{a} & $\Lambda$CDM & $\Lambda$CDM & $\Lambda$CDM & $\Lambda$CDM \\
 &&  & + $dn_s/d\ln k$ & + $r$ & + $N_{\rm eff}$\\
\hline
\hline
Primary&$100\Omega_b h^2$ 
& \ensuremath{2.222\pm0.047} 
& \ensuremath{2.206\pm 0.047} 
& \ensuremath{2.237\pm 0.048}
& \ensuremath{2.238\pm 0.046} 
\\
\LCDM&$\Omega_c h^2$ 
& \ensuremath{0.113\pm0.0034} 
& \ensuremath{0.1148\pm 0.0039} 
& \ensuremath{0.1117\pm 0.0033}
& \ensuremath{0.140\pm 0.015} 
\\
&$\Omega_\Lambda$ 
& \ensuremath{0.724\pm0.016} 
& \ensuremath{0.713\pm 0.019} 
& \ensuremath{0.729\pm 0.017}
& \ensuremath{0.715\pm 0.017} 
\\
&$n_s$ 
& \ensuremath{0.963\pm0.011} 
& \ensuremath{1.017\pm0.036} 
& \ensuremath{0.970\pm0.012}
& \ensuremath{0.983\pm0.014} 
\\
&$\tau$ 
& \ensuremath{0.086\pm0.013} 
& \ensuremath{0.095\pm 0.016} 
& \ensuremath{0.086\pm 0.015}
& \ensuremath{0.086\pm 0.014} 
\\
&$10^9\Delta_{\cal R}^2$ 
& \ensuremath{2.46\pm0.09} 
& \ensuremath{2.39\pm 0.10} 
& \ensuremath{2.40\pm 0.10}
& \ensuremath{2.44\pm 0.09}
\\
\hline
Extended&$dn_s/d\ln k$ & &\ensuremath{-0.024\pm 0.015} &
\\
&$r$ & & &\ensuremath{<0.19}&
\\
&$N_{\rm eff}$ & && &\ensuremath{4.56\pm 0.75}
\\
\hline
Derived &$\sigma_8$ 
& \ensuremath{0.813\pm0.022} 
& \ensuremath{0.820\pm 0.023} 
& \ensuremath{0.811\pm 0.022}
& \ensuremath{0.885\pm 0.039} 
\\
&$\Omega_m$ 
& \ensuremath{0.276\pm0.016} 
& \ensuremath{0.287\pm 0.019} 
& \ensuremath{0.271\pm 0.017}
& \ensuremath{0.285\pm 0.017}
\\
&$H_0$ 
& \ensuremath{69.9\pm1.4} 
& \ensuremath{69.1\pm1.5} 
& \ensuremath{70.4\pm1.5}
& \ensuremath{75.5\pm3.0} 
\\
\hline
Secondary &$B^{\rm SZ}_{3000}$~$(\mu {\rm K}^2)$ 
& \ensuremath{<9.7} 
& \ensuremath{<11.4} 
& \ensuremath{<10.2}
& \ensuremath{<12.1}
\\
&$A_p$~$(\mu {\rm K}^2)$ 
& \ensuremath{16.1\pm2.0} 
& \ensuremath{15.2\pm 2.0} 
& \ensuremath{16.1\pm2.0}
& \ensuremath{15.3\pm 2.1} 
\\
&$A_c$~$(\mu {\rm K}^2)$  
& \ensuremath{<8.4} 
& \ensuremath{<10.3} 
& \ensuremath{<8.4}
& \ensuremath{<10.2}
\\
\hline
\hline
\tablenotetext{a}{For one-tailed distributions, the upper 95\% CL is given. For two-tailed distributions the 68\% CL are shown.}
\end{tabular}
\label{table:lcdm+all_params}
\end{center}
\end{table*}

\begin{figure}[t]
  \centering
  \resizebox{.5\textwidth}{!}{
  \plotone{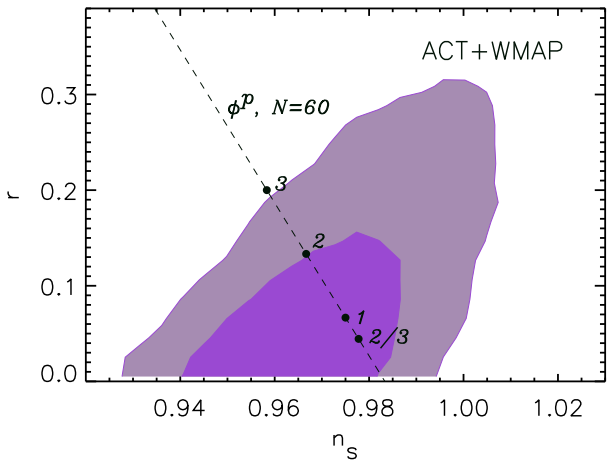} 
} 
\caption{Two-dimensional marginalized distribution (68\% and 95\% CL) for the tensor-to-scalar ratio $r$, and the scalar spectral index $n_s$, for \act+\wmap\ data. By measuring the $\ell>1000$ spectrum, the longer lever arm from \act\ data further breaks the $n_s-r$ degeneracy,  giving a marginalized limit $r<0.25$ (95\% CL) from the CMB alone. The predicted values for a chaotic inflationary model with inflaton potential $V(\phi)\propto \phi^p$ with 60 e-folds are shown for $p=3,2,1,2/3$; $p>3$ is disfavored at $>95$\% CL.}
  \label{fig:tensors}
\end{figure}

\begin{figure*}[htb]
  \centering
  \resizebox{.9\textwidth}{!}{
  \plotone{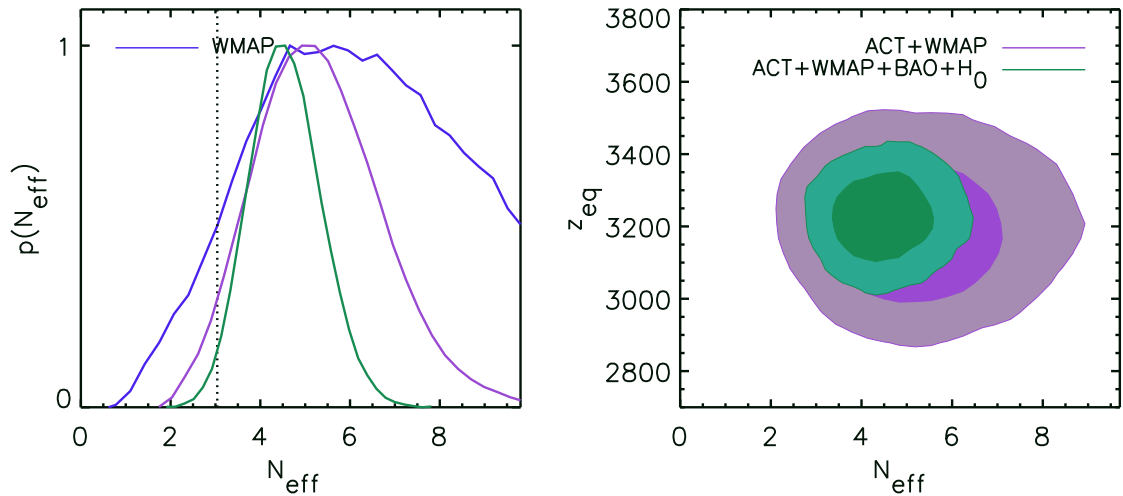} 
} 
\caption{Constraints on the effective number of relativistic species, \neff. {\it Left}: One-dimensional marginalized distribution for \neff, for data combinations indicated in the right panel. The standard model assumes three light neutrino species (\neff=3.04, dotted line); the mean value is higher, but 3.04 is within the 95\% CL. {\it Right:} Two-dimensional marginalized distribution for \neff\ and equality redshift $z_{\rm eq}$, showing that \neff\ can be measured separately from $z_{\rm eq}$. \neff\ is bounded from above and below by combining the small-scale \act\ measurements of the acoustic peaks with \wmap\ measurements.  The limit is further tightened 
by adding BAO and $H_0$ constraints, breaking the degeneracy between \neff\ and the matter density  by measuring the expansion rate at late times. 
}

  \label{fig:neff}
\end{figure*}

\subsubsection{Gravitational waves}
\label{subsubsec:grav_waves}
The concordance \LCDM\ model assumes purely scalar fluctuations. Tensor fluctuations can also be seeded at early times, propagating as gravitational waves. They contribute to the CMB temperature and polarization anisotropy, polarizing the CMB with both an E-mode and B-mode pattern \citep[e.g.,][]{kamionkowski/kosowsky/stebbins:1997,zaldarriaga/seljak:1997}. The tensor fluctuation power is quantified using the tensor-to-scalar ratio $r =\Delta_h^2(k_0)/\Delta_{\cal R}^2(k_0)$, where $\Delta_h^2$ is 
the amplitude of primordial gravitational waves, with pivot scale $k_0=0.002/{\rm Mpc}$. Inflationary models predict tensor fluctuations, with amplitude related to the potential of the inflaton field (see e.g., \citet{baumann/etal:2009} for a recent review.)

Direct B-mode polarization measurements from the BICEP experiment provide limits of $r<0.7$ (95\% CL, \citet{chiang/etal:2010}). Temperature and E-mode fluctuations over a range of scales currently provide a stronger indirect constraint on $r$, with \ensuremath{r < 0.36\ \mbox{(95\% CL)}} from the \wmap\ data \citep{komatsu/etal:prep}. Models with a large value for $r$ have increased power at large scales, which can be partly compensated by increasing the spectral index of scalar fluctuations and reducing the scalar amplitude. This `$n_s-r$' degeneracy can be partly broken with lower-redshift observations that limit \ensuremath{r < 0.24\ \mbox{(95\% CL)}} from \wmap+BAO+$H_0$ \citep{komatsu/etal:prep}. It can also be broken by measuring temperature fluctuations at $\ell>1000$. The tightest CMB-only constraints so far have come from \wmap\ CMB data combined with 
ACBAR and QUAD small-scale CMB data \citep{pryke/etal:2008},  with 
\ensuremath{r < 0.3\ \mbox{(95\% CL)}}. With \act\ combined with \wmap\ we now find 
\be
r\ensuremath{<0.25}~(95\%~{\rm CL})
\ee
for the CMB temperature anisotropy power spectrum alone, comparable to constraints from combined cosmological datasets ($r\ensuremath{<0.19}$~ at 95\%~CL for ACT+\wmap+BAO+$H_0$). The parameter estimates are given in Tables \ref{table:lcdm_params} and \ref{table:lcdm+all_params}, and the dependence of the tensor amplitude on the spectral index is shown in Figure \ref{fig:tensors}. For chaotic inflationary models with inflaton potential $V(\phi)\propto \phi^p$ and $N$ e-folds of inflation, the predicted tensor-to-scalar ratio is $r=4p/N$, with $n_s=1-(p+2)/2N$. The CMB data exclude $p \ge 3$ at more than 95\% confidence for $N=60$ e-folds.

\subsection{Non-standard models}
\label{subsec:non_standard}
In addition to specifying the primordial perturbations, the concordance model assumes that there are three light neutrino species, that standard BBN took place with specific predictions for primordial element abundances, and that there are no additional particles or  fluctuations from components such as cosmic defects. The damping tail measured by ACT offers a probe of possible deviations from this standard model. 

\subsubsection{Number of relativistic species}
\label{subsubsec:neff}

\begin{figure*}[t]
  \centering
  \resizebox{.9\textwidth}{!}{
  \plotone{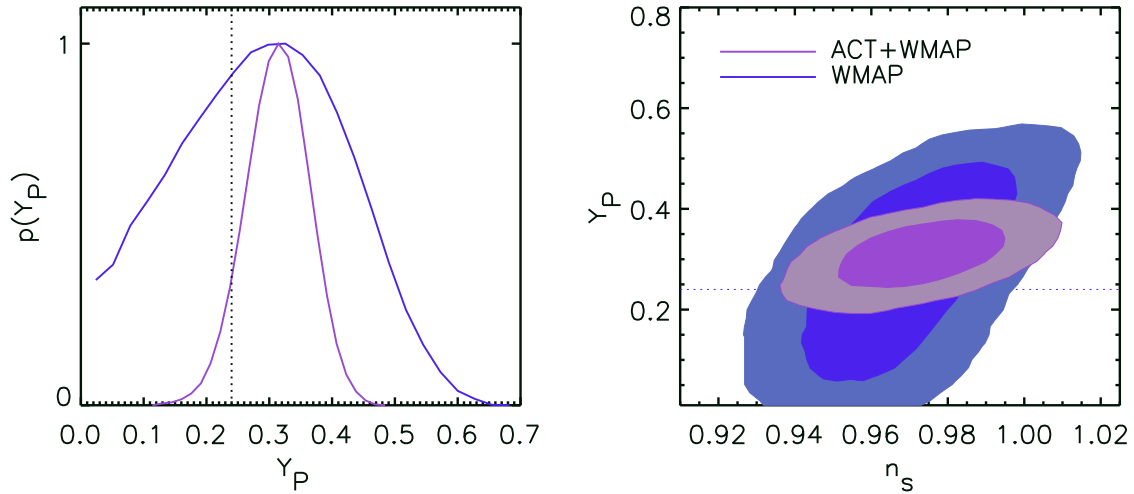} 
} 
\caption{Constraint on the primordial helium mass fraction $Y_P$. {\it Left:} The one-dimensional marginalized distribution for $Y_P$ derived from the \act+\wmap\ data compared to \wmap\ alone. The measurement of the Silk damping tail by \act\ constrains the number of free electrons at recombination, giving a $6\sigma$ detection of primordial helium consistent with the BBN-predicted $Y_P=0.25$. {\it Right:} The two-dimensional marginalized distribution (68\% and 95\% CL) for $Y_P$ and the spectral index $n_s$; the degeneracy is partly broken with the ACT data.}
  \label{fig:helium}
\end{figure*}

The CMB is sensitive to the number of relativistic species at decoupling.
Changing the effective number of species affects the evolution of 
perturbations by altering the expansion rate of the universe. Neutrinos
also stream relativistically 
out of density fluctuations, with additional species suppressing the CMB peak heights and shifting the acoustic peak positions \citep{ma/bertschinger:1995,bashinsky/seljak:2004}. 

The standard model of particle physics has three light neutrino species, consistent with measurements of the width of the Z boson, giving $N_\nu = 2.984 \pm 0.008$ (Particle Data Book). Three neutrino species contribute about 11\% of the energy density of the universe at $z \approx 1100$, with $\rho_{rel} = [7/8(4/11)^{4/3}N_{\rm eff}]\rho_\gamma$. Cosmological datasets are sensitive to  $\rho_{rel}$, which can be composed of any light particles produced during the Big Bang that do not couple to electrons, ions, or photons; or any additional contribution to the energy density of the universe such as gravitational waves. 
Three light neutrino species correspond to \neff=3.04. Any deviation would indicate either additional relativistic species, or evidence for non-standard interactions or non-thermal decoupling \citep{bashinsky/seljak:2004}.

 Recent constraints on the number of relativistic species have been explored with CMB data from \wmap\ combined with low redshift probes by e.g., \citet{spergel/etal:2007,ichikawa/kawasaki/takahashi:2007,mangano/etal:2007,hamann/etal:2007,dunkley/etal:2009,komatsu/etal:prep,reid/etal:2010}. With \wmap\ data a detection was made of relativistic species with \neff$>2.7$ (95\% CL), but the upper level was unconstrained. By combining with distance measures, \citet{komatsu/etal:prep} limit the number of species to \neff$=4.34 \pm0.88$, and \citet{reid/etal:2010} added optical cluster limits and LRG power spectrum measures to find \neff$=3.77\pm0.67$. \citet*{mantz/allen/rapetti:2010} include X-ray cluster gas fraction and cluster luminosity measurements from ROSAT and Chandra to estimate \neff$=3.4^{+0.6}_{-0.5}$, improving limits by constraining the matter power spectrum at low redshift. BBN observations limit $N_{\rm eff}$ to $3.24 \pm 0.6$ \citep{cyburt/etal:2005}. 

By combining the \act\ power spectrum measurement  with \wmap, the effective number of species is estimated from the CMB to be
\be
N_{\rm eff} =  \ensuremath{5.3\pm 1.3}~(68\%~{\rm CL}).
\ee
A universe with no neutrinos is excluded at  4$\sigma$ from the CMB alone, with the marginalized distribution shown in Figure \ref{fig:neff}. We can now put an upper bound on \neff\ from the CMB alone using \act. This improved measurement comes from the third to seventh peak positions and heights. The right panel of Figure \ref{fig:neff} shows the redshift of equality, $z_{\rm eq}$, as a function of the number of species. The relation of $z_{\rm eq}$ to the number of species is given in Eq.~53 of \citet{komatsu/etal:prep}. With large scale measurements the observable quantity from the third peak height is just $z_{\rm eq}$, leading to a strong degeneracy between \neff\ and $\Omega_ch^2$. With small-scale information the CMB data allow a measure of \neff\ in addition to $z_{\rm eq}$ due to the additional effects of anisotropic stress on the perturbations. As an example, a model with \neff$=10$ that fits the \wmap\ data is shown in Figure \ref{fig:l4_spectrum}. With a large $N_{\rm eff}$ the higher peaks are damped, and slightly shifted to larger multipoles. The model is excluded by the \act\ spectrum in the $1000<\ell<2500$ regime. 

The central value for \neff\ preferred by the \act+\wmap\ data is $1.7\sigma$ above the concordance value, with increased damping over the \LCDM\ model; improved measurements of the spectrum will help refine this measurement. This is not interpreted as a statistically significant departure from the concordance value; the best-fit $\chi^2$ is only 1.3 less than for \neff=3.04. The degeneracy between \neff\ and $\Omega_ch^2$ results in a higher mean value for $\sigma_8$, \ensuremath{0.906\pm 0.059}, with all estimated parameters given in Table \ref{table:lcdm_params}. By adding the BAO and $H_0$ data the \neff$-\Omega_ch^2$ degeneracy is further broken, with $N_{\rm eff} =   \ensuremath{4.56\pm 0.75}$ (68\% CL). 
This central value is higher than from joint constraints including X-ray and optical cluster measurements \citep{reid/etal:2010,mantz/allen/rapetti:2010}; improved CMB and low redshift measurements will allow further constraints and consistency checks.

\subsubsection{Primordial $^4$He abundance}
\label{subsubsec:helium}

In the standard BBN model, light nuclides are synthesized in the first few minutes after the Big Bang. Measurements of the abundance of helium are therefore sensitive to the expansion rate of the universe during this time \citep{peebles:1966,steigman/schramm/gunn:1977}. In standard BBN, the expected $^4$He mass fraction, $Y_P$, is related to the baryon density, $\Omega_bh^2$, and the number of neutrino (or relativistic) species, $N_{\rm eff}$, by
\be
Y_P = 0.2485 + 0.0016[(273.9 \Omega_bh^2 -6) + 100(S-1)],
\label{eqn:bbn}
\ee
where $S^2= 1+(7/43)(N_{\rm eff}-3)$ \citep[see e.g.,][]{kneller/steigman:2004,steigman:2007,simha/steigman:2008}. For the $\Lambda$CDM model, with baryon density $100\Omega_bh^2 = 2.214\pm0.050$ and $N=3.04$ effective species, the predicted helium fraction is $Y_P=0.2486 \pm0.0006$, with error dominated by the 0.02\% uncertainty on the linear fit in Eq.~\ref{eqn:bbn} \citep{steigman:2010}. When the neutrino species are allowed to vary (as in Section \ref{subsubsec:neff}), the current prediction from \act+\wmap+BAO+$H_0$ is $Y_P=0.267\pm0.009$. For comparison, the prediction from the baryon density derived from deuterium abundance measurements is $Y_P = 0.2482\pm0.0007$ \citep[see][for a review]{steigman:2010}. A measurement of any deviation from this prediction could point the way to non-standard models, in particular those that affect the timing of BBN \citep{steigman/schramm/gunn:1977, boesgaard/steigman:1985,jedamzik/pospelov:2009}. This includes modifications to the Hubble expansion rate during BBN, energy injection due to annihilation or decay of heavy particles, particle catalysis of BBN reactions, and time variations in fundamental constants (see, e.g., \citet{peimbert:2008,jedamzik/pospelov:2009} for discussions). 

The $^4$He abundance estimated from observations of metal poor extragalactic regions \citep[see][for example]{steigman:2007,peimbert:2008} is $Y_P=0.252\pm0.004$ and $0.252\pm0.001$ \citep{izotov/thuan/stasinska:2007}, although a higher measurement of $Y_P=0.2565\pm0.0010({\rm stat})\pm0.0050({\rm syst})$ has recently been made \citep{izotov/thuan:2010}. There are systematic uncertainties in the astrophysically derived abundances, as helium is depleted in stars.  

The CMB provides an alternative probe of the helium abundance when the universe was $\simeq$400,000 years old. Helium recombines earlier than hydrogen, at $z\approx1800$ rather than $z\approx 1100$, reducing the number density of electrons at recombination to $n_e=n_b(1-Y_P)$, where $n_b$ is the baryon number density \citep{hu/etal:1995}. It affects the CMB at small scales by modifying the recombination process. A larger $Y_P$ decreases the electron number density, increasing the mean free path of Compton scattering. This leads to decreased power on small scales, due to enhanced Silk damping, as shown in \citet{trotta/hansen:2004} and \citet{komatsu/etal:prep}.

\begin{figure}[t]
  \centering
  \resizebox{.5\textwidth}{!}{
  \plotone{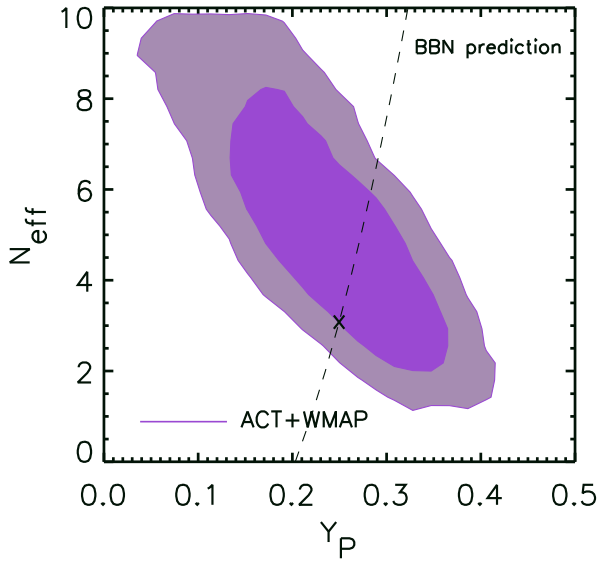} 
} 
\caption{Joint two-dimensional marginalized distribution (68\% and 95\% CL) for the primordial helium mass fraction $Y_P$ and the number of relativistic species \neff. The two are partly degenerate, as increasing \neff\ or $Y_P$ leads to increased damping of the power spectrum. The predicted standard-BBN relation between \neff\ and $Y_P$ is indicated. The concordance \neff=3.04, $Y_P=0.25$ model lies on the edge of the two-dimensional 68\% CL, and a model with \neff=0, $Y_P=0$ is excluded at high significance.}
\vskip +0.2cm
  \label{fig:neff_yhe}
\end{figure}

For CMB analysis the primordial helium abundance is usually assumed to be  $Y_P=0.24$. Constraints on a varying abundance from the CMB have been presented in \citet{trotta/hansen:2004,huey/cyburt/wandelt:2004,ichikawa/takahashi:2006,ichikawa/sekiguchi/takahashi:2008,dunkley/etal:2009,komatsu/etal:prep}, with a $>3\sigma$ detection of $Y_P =0.33\pm0.08$ reported in \citet{komatsu/etal:prep} for the seven-year \wmap\ data combined with small-scale CMB observations from ACBAR \citep{reichardt/etal:2009} and QUAD \citep{pryke/etal:2008}. 
We now find 
\be
Y_P=\ensuremath{0.313\pm 0.044}~(68\%~{\rm CL})
\ee
 with the \act+\wmap\ data combination, a significant detection of primordial helium from the CMB alone. The mean value is higher than predicted from \LCDM, but consistent at the 1.5$\sigma$ level. A universe with no primordial helium is ruled out at $6\sigma$. The distributions for $Y_P$ and its correlation with the spectral index are shown in Figure \ref{fig:helium}, with statistics in Table \ref{table:lcdm_params}. Figure \ref{fig:l4_spectrum} shows how a higher helium fraction consistent with \wmap\ data ($Y_P=0.5$) is ruled out by ACT's determination of Silk damping at small scales.  There is still some uncertainty in the exact details of recombination \citep[e.g.,][]{wong/scott:2007,chluba/rubino-martin/sunyaev:2007,switzer/hirata:2008,fendt/etal:2009,chluba/sunyaev:2010}. A recent refinement of the numerical code for recombination used for this analysis (Recfast 1.5, by \citet{seager/sasselov/scott:1999}, updated to match \citet{rubino-martin/etal:2010}), gives a $2\%$ change in the spectrum at $\ell=2000$. This is subdominant to the $7\%$ percent shift from a $1\sigma$ change in $Y_P$, so these effects are not expected to significantly affect current constraints, although will become more important as the data improve. 
If we consider the possible variation of both the primordial helium fraction and the number of relativistic species, the constraints on each parameter are weakened as there is some degeneracy between the two effects. The joint marginalized distribution for these parameters is shown in Figure \ref{fig:neff_yhe}, together with the predicted relation between \neff\ and $Y_P$ assuming standard BBN. The concordance \neff=3.04, $Y_P=0.25$ model lies on the edge of the two-dimensional 68\% CL, and a model with \neff=0, $Y_P=0$, is excluded at high significance.

\subsubsection{Cosmic strings}
\label{subsubsec:strings}

\begin{figure}[t]
  \centering
\hskip -0.4cm
  \resizebox{.5\textwidth}{!}{
  \plotone{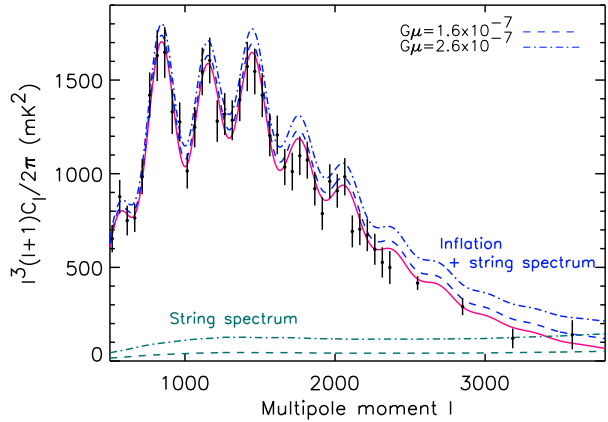} 
} 
\caption{The power spectrum measured by \act\ at \arone, scaled by $\ell^4$ and with best-fit secondary model subtracted, with the best-fit \LCDM\ (solid) compared to a model with maximal cosmic string tension $G\mu=1.6\times 10^{-7}$ allowed by the ACT data at 95\% CL (dashed, assuming a Nambu string template described in Section \ref{subsubsec:strings}). A model with the 95\% upper limit allowed without including ACT data, with $G\mu=2.6\times 10^{-7}$, is shown for comparison (dot-dashed); it overpredicts the observed power in the range $1500<\ell<4000$.}
  \label{fig:string}
\end{figure}

 Observations of the acoustic peaks in the CMB power spectrum have ruled out defects from phase transitions as the dominant mechanism for seeding cosmic structure \citep[see e.g.,][]{vilenkin/shellard:2000}. However, certain inflation models predict string perturbations of similar amplitudes to the inflationary perturbations \citep{linde:1994,dvali/tye:1999}. Using the CMB one can constrain the string tension, $G\mu$, and therefore the energy scale at which the strings are formed. Unfortunately there is significant uncertainty in the predicted power spectrum from cosmic string-generated anisotropies, due to difficulties in modeling the string network. Most approaches model the network as an ensemble of string segments with constant average properties, with string loops produced that decay into radiation. The equations of motion are solved using either the Nambu or Abelian-Higgs (AH) method, described in e.g., \citet{bennett/bouchet:1990,pogosian/vachaspati:1999,bevis/etal:2007,battye/moss:prep}. 

The small-scale CMB provides a unique probe of cosmic strings, with simulations and forecasts by \citet{fraisse/etal:2008} and \citet{bevis/etal:prep} predicting a power law behavior that could dominate over the Silk damping tail of the inflationary inhomogeneities, consistent with analytic predictions by \citet{hindmarsh:1994}.
Constraints have been placed on the cosmic string tension from recent CMB and other cosmological data \citep{lo/wright:2005,wyman/pogosian/wasserman:2005,wyman/pogosian/wasserman:2006,battye/garbrecht/moss:2006,fraisse:2007,bevis/etal:2007,urrestilla/etal:2008,sievers/etal:2009}. Most recently, \citet{battye/moss:prep} report limits of $G\mu < 2.6\times 10^{-7}$ (95\% CL) for Nambu strings using 5-year WMAP data combined with large-scale structure and BBN data. They find a significant dependence of this limit on the chosen string model, with up to a factor of three variation. For a simple comparison, we consider just the Nambu cosmic string template used in \citet{battye/moss:prep}, extended to scales $\ell>3000$ with a power law, ${\cal B}_\ell \propto \ell^{-1}$ \citep{fraisse/etal:2008}. The template is held fixed for all cosmological models. Assuming this model, we find limits from ACT combined with \wmap\ of $q_{\rm str}< 0.025$ (95\% CL), which corresponds to a tension of 
\be
G\mu < 1.6\times 10^{-7}~(95\%~{\rm CL}).
\ee
The spectrum corresponding to this 95\%  upper limit is shown in Figure \ref{fig:string}, compared to the upper limit pre-ACT given in \citet{battye/moss:prep} which overpredicts the power measured by ACT in the range $1500<\ell<4000$. 
The joint constraint on the string tension and the scalar spectral index may also limit the class of hybrid inflation models that produce cosmic strings at the end of inflation, and which typically predict a unity scalar spectral index. \citet{battye/garbrecht/moss:2006} and \citet{bevis/etal:2008} demonstrated that these models provided a good fit to the data, but more recently \citet*{battye/garbrecht/moss:2010} found that minimal D-term models are now ruled out at a 4$\sigma$ level with CMB combined with SDSS and BBN data, and that minimal F-term models are increasingly disfavored. The preference from CMB data alone for a red spectrum with $n_s=\ensuremath{0.963\pm0.013}$ (ACT+\wmap), marginalized over a string contribution, provides further evidence against these hybrid models.

\section{Discussion}
\label{sec:discuss}

The power spectra measured at \arone\ and \artwo\ by ACT, using observations made in the Southern sky in 2008, have provided a new probe of the physics affecting microwave fluctuations at small scales. The concordance \LCDM\ cosmological model continues to be favored, and possible deviations from this model are more tightly constrained. At \arone\ and \artwo, the CMB is dominant at scales larger than $\ell\simeq3000$ and $2000$ respectively, after bright sources have been removed. At smaller scales, a simple model for SZ and point source emission is a good fit to the \act\ power spectra. 
By using multi-frequency information we see a preference for non-zero thermal Sunyaev-Zel'dovich fluctuations in the \arone\ power spectrum at 2$\sigma$, with an amplitude consistent with observations by the South Pole Telescope. The mean amplitude is lower than the simplest cluster models predict for a universe with $\sigma_8=0.8$, but at less than 2$\sigma$ significance. The level is consistent with expectations from recent models that include more complex gas physics; continued comparisons of observations and theory will allow progress to be made on cluster modeling.  The frequency dependence of the infrared emission has the expected behavior of graybody emissivity from dusty star-forming galaxies at redshifts $1<z<4$, and a clustered infrared point source component is detected in the ACT data at 5$\sigma$ significance.

The $1000<\ell<3000$ spectrum provides a measure of the third to seventh acoustic peaks in the CMB, and the Silk damping tail from the recombination process at $z=1100$.  Using this measurement we place tighter constraints on deviations from the \LCDM\ model. Given the uncertainty on the expected power spectrum from infrared point sources, we have limited this part of the analysis to the power spectrum from \arone. The data are found to be consistent with no deviations from \LCDM, and the gravitational lensing of the temperature power spectrum is at the expected level, with an unlensed signal disfavored at $2.8\sigma$.  We have detected primordial helium at 6$\sigma$, and relativistic species at 4$\sigma$ from the CMB alone, both consistent with the expected levels. The best-fit models prefer increased damping beyond the \LCDM\ expectation, but at less than the 2$\sigma$ level. The cosmological parameters considered have distinct effects on the power spectrum, but there is some degeneracy between \neff, $Y_P$, and $dn/d \ln k$. This means that an enhanced damping leads to higher mean values for either \neff\ or $Y_p$, or to more negative $dn/d \ln k$.  These are seen at 1.4-1.8$\sigma$ from the concordance value when considered individually as extensions to \LCDM, but there is no evidence that these deviations are statistically preferred. We do not find evidence for a gravitational wave component or a contribution from cosmic string fluctuations, indicating the continued consistency of cosmological data with minimal inflationary models.

\acknowledgements

ACT is on the
Chajnantor Science preserve, which was made possible by the Chilean
Comisi\'on Nacional de Investigaci\'on Cient\'ifica y Tecnol\'ogica.
We are grateful for the assistance we received at various times from
the ALMA, APEX, ASTE, CBI/QUIET, and NANTEN2 groups.  
The PWV data come from the public APEX weather website.
Field operations were based at the Don Esteban facility run by
Astro-Norte. Reed Plimpton and David Jacobson
worked at the telescope during the 2008 season. 
We thank Norm Jarosik for support throughout the project.  
We also thank 
Adam Moss and Richard 
Battye for sharing their cosmic string power spectrum, Laurie Shaw 
and Nick Battaglia for providing SZ power spectra, and Bruce Bassett for 
suggestions on testing lensing in the power spectrum. We thank Marco Viero and Graeme Addison for providing useful input on clustered point sources.
This work was supported by the U.S. National Science Foundation
through awards AST-0408698 for the ACT project, and PHY-0355328,
AST-0707731 and PIRE-0507768. Funding was also provided by Princeton
University and the University of Pennsylvania.  The PIRE program made
possible exchanges between Chile, South Africa, Spain and the US that
enabled this research program.  
JD acknowledges support from an RCUK Fellowship.  RH
received funding from the Rhodes Trust. VA, SD, AH, and TM were supported through NASA grant NNX08AH30G.  ADH
received additional support from a Natural Science and Engineering
Research Council of Canada (NSERC) PGS-D scholarship. AK and BP were
partially supported through NSF AST-0546035 and AST-0606975,
respectively, for work on ACT\@.  LI acknowledges partial support
from FONDAP Centro de Astrof\'isica.  RD was supported by CONICYT,
MECESUP, and Fundaci\'on Andes.  ES acknowledges support by NSF
Physics Frontier Center grant PHY-0114422 to the Kavli Institute of
Cosmological Physics. KM, M Hilton, and RW received financial support
from the South African National Research Foundation (NRF), the Meraka
Institute via funding for the South African Centre for High
Performance Computing (CHPC), and the South African Square Kilometer
Array (SKA) Project.    SD 
acknowledges support from the Berkeley Center for Cosmological
Physics.  YTL acknowledges support from the World Premier
International Research Center Initiative, MEXT, Japan. NS is supported by the U.S. Department of Energy contract to SLAC no. DE-AC3-76SF00515. 
We acknowledge the use of the Legacy Archive for Microwave Background Data Analysis (LAMBDA). Support for LAMBDA is provided by the NASA Office of Space Science.  The data will be made public through LAMBDA (\url{http://lambda.gsfc.nasa.gov/}) and the ACT website (\url{http://www.physics.princeton.edu/act/}).

\appendix
\label{Multi-frequency likelihood} 

\section{Beam Likelihood}
\label{sec:beam_like}

Following the prescription in \cite{hinshaw/etal:2007}, the likelihood of the 
beam-deconvolved spectrum is written as 
\be
{\cal L} = \sum_{bb'} 
 ({\hat C}_b-C_b) ({\bf Q}_0 + {\bf Q}_1)^{-1}_{bb'} ({\hat C}_{b'}-C_{b'})
 + \ln\det ({\bf Q}_0 + {\bf Q}_1).
\ee
where ${\bf Q}_0$ contains noise and cosmic variance, and ${\bf Q}_1$ contains beam error.  The matrix ${\bf Q}_1$ is diagonalized in the form
\be
{\bf Q}_1 \approx {\bf U U}^T,
\ee
where ${\bf U}$ is an $N_b \times M$ matrix with $M \ll N_b$. 
The matrix ${\bf U}$ is computed by decomposing the unbinned beam covariance matrix {${\bf \Sigma}_b$}, the covariance between normalized {$b_\ell$} and {$b_{\ell'}$}, as
\be
{\bf \Sigma_b} \approx {\bf P D P}^T.
\ee
The matrix ${\bf Q}_1$ is related to the beam covariance matrix {${\bf \Sigma}_b$} by
\be
{\bf Q}^{ll'}_{1} =4 \frac{C_\ell}{b_\ell} {\bf \Sigma_b}\frac{C_\ell}{b_\ell},
\ee
so the elements of matrix ${\bf U}$ are given by 
\be
U_{b,i} = M_{b\ell}[2 \frac{C_\ell}{b_\ell} P_{\ell,i} \sqrt{D_i}], 
\ee
where $M_{bl}$ are the bandpower window functions.  Once decomposed, the Woodbury formula gives
\ba
({\bf Q}_0 + {\bf Q}_1)^{-1} & \approx & ({\bf Q}_0 + {\bf U U}^T)^{-1} \\
                 & = & {\bf Q}_0^{-1} - {\bf Q}_0^{-1} {\bf U} 
      \left({\bf I} + {\bf U}^T {\bf Q}_0^{-1} {\bf U} \right)^{-1} 
      {\bf U}^T {\bf Q}_0^{-1}.
\ea
The likelihood is then given by ${\cal L} = {\cal L}_0 + {\cal L}_b$, where
\be
{\cal L}_0 = \sum_{bb'} 
 ({\hat C}_b-C_b) ({\bf Q}_0)^{-1}_{bb'} ({\hat C}_{b'}-C_{b'})
 + \ln\det {\bf Q}_0,
\ee
and
\be
{\cal L}_b = -\sum_{bb'} ({\hat C}_b-C_b) M^{-1}({\hat C}_{b'}-C_{b'}) + N_M 
\ee
where
\be
M^{-1}= \left[{\bf Q}_0^{-1} {\bf U} 
      \left({\bf I} + {\bf U}^T {\bf Q}_0^{-1} {\bf U} \right)^{-1} 
      {\bf U}^T {\bf Q}_0^{-1} \right]_{bb'} 
\ee
and $N_M=\ln\det({\bf I} + {\bf U}^T {\bf Q}_0^{-1}{\bf U})$.

\section{Sensitivity to likelihood assumptions}
\label{sec:like_sensit}

 A set of assumptions are made in the \act\ likelihood. We choose the \LCDM+running model to check their effect on cosmological parameters, as subtleties in the small-scale treatment can be probed more thoroughly with this model than with the 6-parameter \LCDM.  The fiducial constraints on the \LCDM+running model use data between $500<\ell<10000$, including beam error and a $2\%$ calibration error in temperature, and use the Src-1 clustered source template and the TBO-1 SZ template. The assumptions we test are the choice of SZ and clustered source templates, the dependence of results on the beam and calibration error, and the range of angular scales used.

\begin{figure}[htb]
  \centering
  \resizebox{.8\textwidth}{!}{
  \plotone{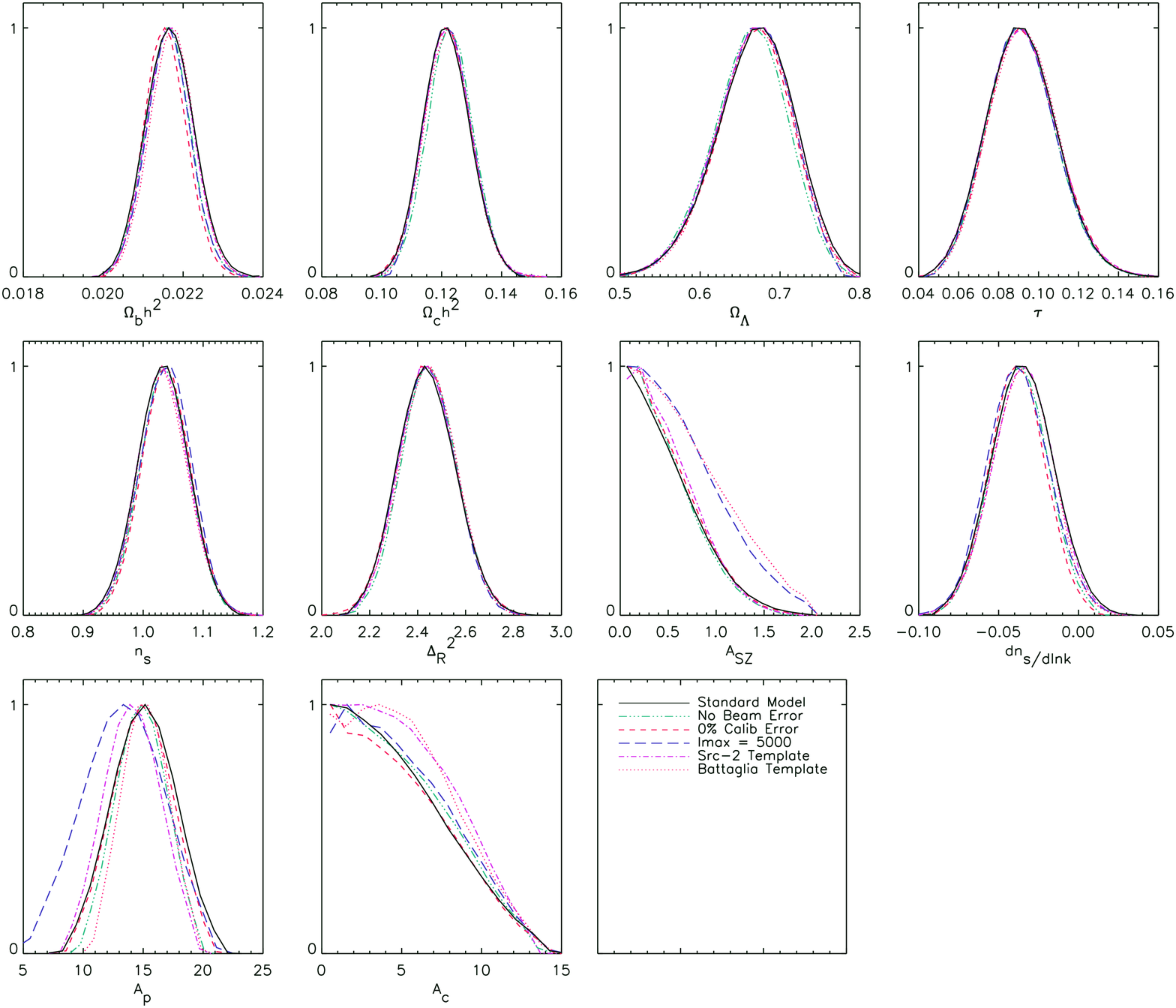} 
} 
\caption{Tests of the \arone\ likelihood for the \LCDM\ model with a running spectral index. The `standard' model uses the likelihood settings described in Sec \ref{subsec:act_like}. The primary cosmological parameter constraints are not affected if we assume different template spectra for the clustered point source component (`Src-2') and for the SZ spectrum (`Battaglia'). Changing the angular range used from $\ell_{max}=10000$ to $5000$  only alters the distribution of the secondary parameters, but leaves the primary model parameters unchanged. Parameters are tightened by about $0.1\sigma$ if the beam error is set to zero (`No beam error'), and by up to $0.2\sigma$ is the calibration error is removed (`0\% Calib error').} 
  \label{fig:like_test}
\end{figure}

Substituting the alternative halo-model clustered source template, Src-2, has a neglible effect on the primary cosmological parameters, as shown in Figure \ref{fig:like_test}. Similarly, using the Battaglia SZ template \citep{battaglia/etal:prep}, does not affect primary parameters. The distribution of the SZ amplitude is broadened in this case as the template has a lower amplitude (as described in Sec \ref{subsec:sz_limit}), and by setting an upper limit on $A_{\rm SZ}<2$ the distribution for $A_p$ is narrowed. We conclude from these tests that the systematic uncertainty on primary parameters from the model is small compared to the statistical uncertainty. 

We also test the sensitivity to the calibration and beam uncertainty. Removing the 2\% calibration error in the maps decreases the errors on the primary parameters by up to $0.2 \sigma$.  When the beam error is neglected, the constraints are tightened by about $0.1\sigma$, but this is not a significant effect, consistent with the measurement of the beam. If instead the angular range is restricted to $\ell< 5000$, the primary cosmological parameters are unaffected, but the SZ and point source amplitudes are more poorly determined.

\end{document}